\documentclass[aps,prl,onecolumn,superscriptaddress]{revtex4-1}

\usepackage{graphicx}
\usepackage{subfigure}
\usepackage{dcolumn}
\usepackage{bm}
\usepackage{amsmath}
\usepackage{color}
\usepackage{verbatim}
\usepackage{natbib}
\usepackage{verbatim}
\usepackage{multirow}
\usepackage[colorlinks,
            linkcolor=blue,
            anchorcolor=blue,
            citecolor=blue,
urlcolor=blue
            ]{hyperref}
\usepackage{color}
\usepackage{enumerate}
\usepackage{makecell}

\usepackage{titletoc}

\begin{document}

\title{Progress and Perspectives on Weak-value Amplification}


\author{Liang Xu}
\email{liangxu.ceas@nju.edu.cn}
\affiliation{National Laboratory of Solid State Microstructures, Key Laboratory of Intelligent Optical Sensing and Manipulation, College of Engineering and Applied Sciences, and Collaborative Innovation Center of Advanced Microstructures, Nanjing University, Nanjing 210093, China}

\author{Lijian Zhang}
\email{lijian.zhang@nju.edu.cn}
\affiliation{National Laboratory of Solid State Microstructures, Key Laboratory of Intelligent Optical Sensing and Manipulation, College of Engineering and Applied Sciences, and Collaborative Innovation Center of Advanced Microstructures, Nanjing University, Nanjing 210093, China}

\date{\today}
\begin{abstract}
Weak-value amplification (WVA) is a metrological protocol that effectively amplifies ultra-small physical effects, making it highly applicable in the fields of quantum sensing and metrology. However, the amplification effect is achieved through post-selection, which leads to a significant decrease in signal intensity. Consequently, there is a heated debate regarding the trade-off between the amplification effect and the success probability of post-selection, questioning whether WVA surpasses conventional measurement (CM) in terms of measurement precision. Extensive research indicates that the specific theoretical assumptions and experimental conditions play crucial roles in determining the respective advantages of WVA and CM. WVA provides new perspectives for recognizing the important role of post-selection in precision metrology. It demonstrates significant advantages in two aspects: (i) WVA based on the phase space interaction provides feasible strategies to practically achieve the Heisenberg-scaling precision using only classical resources. (ii) WVA exhibits robustness against certain types of technical noise and imperfections of detectors. Moreover, WVA allows for various modifications to extend the applicable scope and enhance the metrological performance in corresponding situations. Despite substantial progress in recent years, the inherent connection between the advantages of WVA and its unique features remains incompletely understood. In this paper, we systematically review the recent advances in the WVA scheme, with a particular focus on the ultimate precision of WVA under diverse conditions. Our objective is to provide a comprehensive perspective on the benefits of WVA in precision measurement and facilitate the realization of its full potential.
\end{abstract}

\maketitle

\begin{titlepage}
\tableofcontents
\end{titlepage}





\section{1. Introduction}
The precise measurement of ultra-small physical parameters has been instrumental in advancing fundamental research \cite{abbott2016observation, baker2006improved, bulatowicz2013laboratory} and driving the revolution of advanced technology \cite{budker2007optical, kornack2005nuclear, budker2007optical}. However, experimental imperfections unavoidably diminish the practical sensitivity and precision of measurement protocols, making it challenging to effectively extract weak signals. To enhance the sensitivity, researchers have pursued two complementary approaches: designing and manufacturing sophisticated equipment to isolate noise sources, and developing noise-robust measurement protocols. Among the most notable noise-robust measurement protocols, weak-value amplification (WVA) has demonstrated remarkable success in detecting ultra-small physical effects and is increasingly indispensable in the fields of quantum sensing and metrology.

The concept of weak value originates from the influential work of Aharonov, Albert and Vaidman (AAV) in 1988 \cite{aharonov1988result}, wherein they extended the framework of Von Neumann measurement (VNM) \cite{mackey2013mathematical}. In the VNM framework, the quantum measurement is described as a coupling process between a quantum system (QS) and a meter state (MS) under an interaction Hamiltonian. The Hamiltonian that describes this interaction typically takes the form of $\hat{H}_{\text{int}} = g \hat{S}\otimes \hat{M}$, where $g$ is the coupling strength, and $\hat{S}$ ($\hat{M}$) is the observable of the QS (MS). After the interaction, the shift of the MS is then proportional to the eigenvalue of the observable $\hat{S}$ multiplied by the coupling strength $g$. Strong measurement is typically employed to accurately identify measurement results, requiring the coupling strength being much larger than the uncertainty of the MS. In this case, the QS inevitably collapses into the corresponding eigenstate. The AAV extension encompasses two main aspects: (i) weak measurement is performed by setting the coupling strength to be much smaller than the uncertainty of the MS; (ii) The QS is post-selected after the weak measurement with only the surviving MS being observed. In this scenario, both the initial (pre-selected) and final (post-selected) states of the QS jointly determine the average shifts of the post-selected MS, which can be quantified by the weak value of the observable.

The pre- and post-selection is fundamentally connected to the two-state vector formalism (TSVF), which provides a time-symmetric description of the QS. In TSVF, the pre-selected state evolves forward from the past to the future while the post-selected state evolves backward from the future to the past \cite{aharonov1964time, aharonov1990properties, aharonov1998two, muga2007time, reznik1995time}. To reveal the properties of such time-symmetric systems, weak measurement is performed to extract partial information about the QS with minimal disturbance. Thus, weak values maintain the time-symmetric formulation and can be obtained by measuring the post-selected meter states. The measurement of weak values opens up new avenues for exploring fundamental quantum phenomena, including Hardy's paradox \cite{aharonov2002revisiting, lundeen2009experimental}, the three-box paradox \cite{resch2004experimental}, the Leggett-Garg inequality \cite{williams2008weak, dressel2011experimental}, quantum contextuality \cite{pusey2014anomalous, hofmann2020contextuality}, the observation of the Bohm trajectory \cite{kocsis2011observing, mahler2016experimental, xiao2017experimental, xiao2019observing} and more \cite{wiseman2002weak, solli2004fast, huang2019simulating, yu2020experimental, cho2019emergence, chen2019experimental}.

The anomalous weak values that lie outside the eigenvalue spectrum or possess complex values demonstrate promising potential for the development of novel quantum technologies. A comprehensive explanation of weak values from a pragmatic standpoint can be found in Ref. \cite{dressel2014colloquium}. As generalized measurement outcomes, weak values can be used for direct measurement of wavefunction \cite{lundeen2011direct}, which also inspires approaches for direct characterization of generalized quantum systems, including various quantum states, quantum processes and quantum measurements \cite{lundeen2012procedure, salvail2013full, malik2014direct, denkmayr2017experimental, calderaro2018direct, kim2018direct, pan2019direct, PhysRevLett.127.180401, chen2021directly, xu2021direct, xu2024resource}. When the pre- and post-selected states are nearly orthogonal, the large weak value serves as an amplification factor for the coupling strength, giving rise to the WVA scheme. On one hand, this amplification effect helps overcome experimental imperfections and technical noise, contributing to the remarkable achievements of WVA in measuring ultra-small physical parameters. Owing to its advantages, weak value amplification is extensively employed for highly sensitive sensing applications \cite{li2021quantification, qu2020sub, luo2020low, li2018high, cao2024spin}. On the other hand, the amplified outcomes are counterbalanced by a reduction in signal intensity due to post-selection, leading to a potential confusion where WVA may be perceived as suboptimal compared to the CM. The resolution of this contradiction necessitates a detailed evaluation of the ultimate precision achieved by both WVA and CM under various conditions. 

This review employs universal metrics, such as the Fisher information (FI) and the signal-to-noise ratio (SNR) in quantum metrology to evaluate both the performance of WVA and CM. The analysis reveals the distribution of metrological information before and after the post-selection. The optimal WVA allows to achieve almost the same precision as the CM by detecting only a small fraction of successfully post-selected signals. Moreover, the analytical results provide WVA with possible approaches for surpassing the classical limit with only classical resources. The comparison between WVA and CM enhances the understanding of the metrological advantages of WVA under certain conditions. The analytical procedure described here is generally applicable to emerging modified WVA schemes and other metrological protocols that involve post-selection. We hope that this review provides a clear perspective on how the intrinsic properties of WVA contribute to its metrological advantages and inspires the development of noise-robust measurement schemes.
 
The structure of this paper is as follows: section 2 provides an introduction to the theoretical framework of the standard WVA and its generalization. It also presents a comprehensive review of the typical applications of WVA based on different physical implementations. In section 3, we provide a brief overview of the basic framework of quantum metrology, which includes the universal metrics and key precision limits. These metrics are subsequently employed to evaluate the ultimate precision of WVA. Section 4 summarizes various WVA schemes that can achieve Heisenberg-scaling precision, with or without the utilization of quantum resources. Section 5 examines the ultimate precision of WVA in the presence of experimental imperfections, illustrating the advantages of WVA over CM under specific conditions. In section 6, we introduce several modified WVA schemes and highlight their distinct advantages compared to the standard WVA. Finally, the last section concludes our discussion on the advancements made in WVA and proposes potential avenues for future research.

\section{2. Basic principles of Weak-value Amplification}
\subsection{2.1 Standard weak-value amplification}
\label{Sec:weak measurement}
\begin{figure*}[t]
	\centering
	\includegraphics[width=0.9\textwidth]{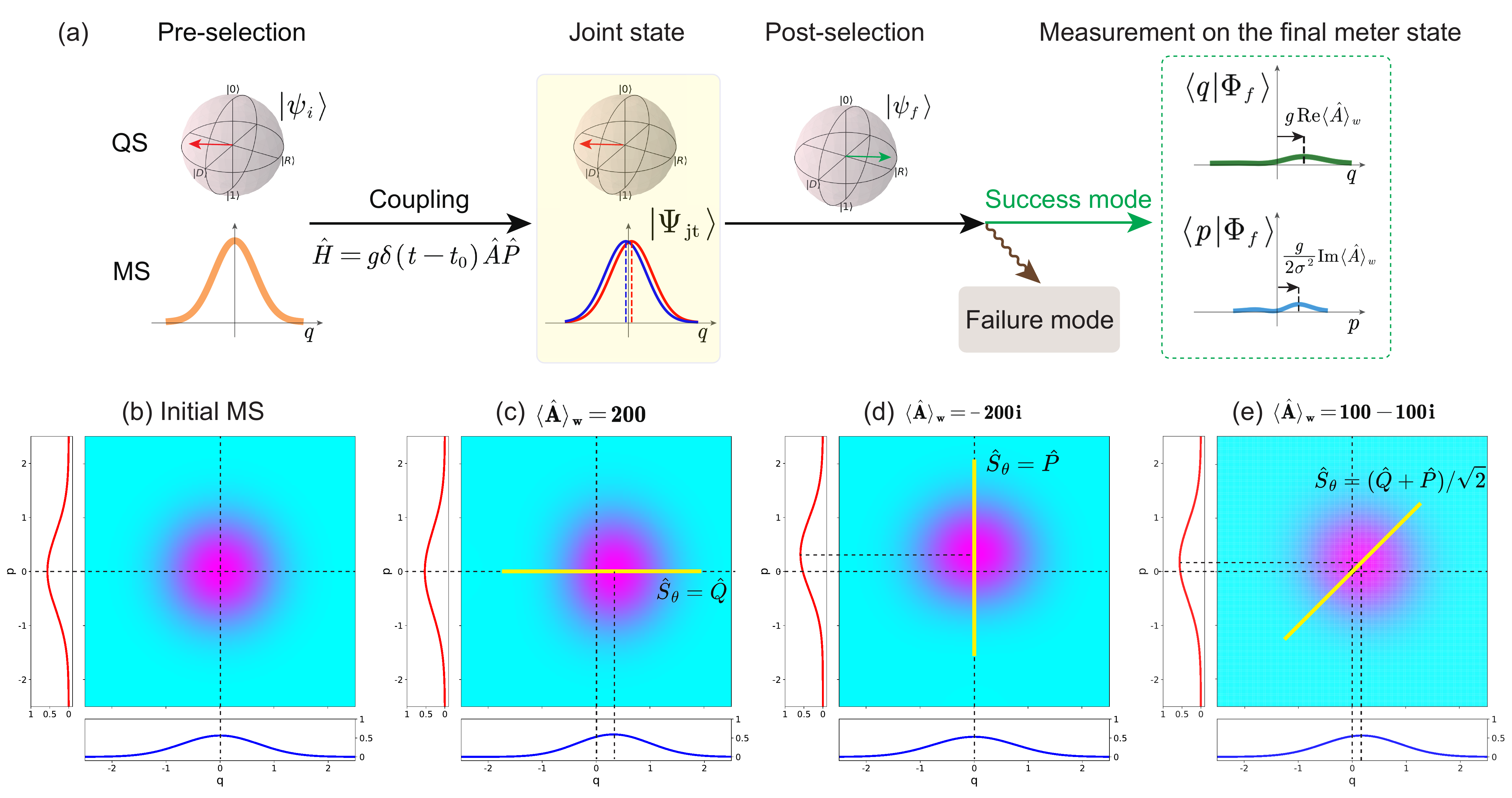}
	\caption{(a) The basic procedure of the standard weak-value amplification scheme. Here, the observable of the QS is set to be $\hat{A} = \hat{\sigma}_z = |0\rangle\langle 0| - |1\rangle\langle 1|$. (b)-(e) The Wigner function and the marginal distribution of the initial meter state and the post-selected meter states with different weak values. Here, we set the parameter of the initial MS $\sigma = \sqrt{2}/2$ and the corresponding parameters of the QS $\langle \hat{A}\rangle_w = 200$: $\theta_i = \pi/2$, $\theta_f = -\pi/2+\epsilon$, $\phi_f-\phi_i=0$; $\langle \hat{A}\rangle_w = -200i$: $\theta_i = \pi/2$, $\theta_f = -\pi/2$,$\phi_f-\phi_i=\phi$; $\langle \hat{A}\rangle_w = 100-100i$: $\theta_i = \pi/2$, $\theta_f = -\pi/2+\epsilon$,$\phi_f-\phi_i=\phi$ with $\epsilon = \phi=0.01$ and $g=\sigma/400$. The yellow lines refer to the optimal observables $\hat{S}_\theta$ in different situations.
}
\label{fig:stdWVA}
\end{figure*}

The typical procedure of WVA is depicted in Fig. \ref{fig:stdWVA} (a). A two-level QS is first pre-selected by $|\psi_i\rangle = \cos{(\theta_i/2)}|0\rangle + \sin(\theta_i/2)e^{i\phi_i}|1\rangle$. The initial MS is prepared as a Gaussian superposition state given by
\begin{eqnarray}
|\Phi\rangle &=& \int dq \frac{1}{(2\pi\sigma^2)^{1/4}}\exp[-\frac{q^2}{4\sigma^2}]|q\rangle {}\nonumber\\
&=& \int dp \Big(\frac{2\sigma^2}{\pi}\Big)^{\frac{1}{4}}\exp(-\sigma^2 p^2)|p\rangle,
\label{eq:initial_MS}
\end{eqnarray}
where $q$ and $p$ are conjugate variables (e.g., momentum and position, respectively). Correspondingly, $|q\rangle$ ($|p\rangle$) denotes the eigenstate of the observable $\hat{Q}$ ($\hat{P}$) of the MS. The QS interacts with the MS through the Hamiltonian $\hat{H}=g\delta(t-t_0)\hat{A}\otimes \hat{P}$, where $\hat{A}$ represents an observable of the QS and $g$ is the coupling strength that needs to be estimated. This interaction leads to the entangled joint state 
\begin{equation}
|\Psi_{jt}\rangle = \exp(-i\int \hat{H}dt)|\psi_i\rangle\otimes |\Phi\rangle.
\label{eq:joint_state}
\end{equation}
Next, the joint state is post-selected by $|\psi_f\rangle =\cos{(\theta_f/2)}|0\rangle + \sin(\theta_f/2)e^{i\phi_f}|1\rangle$ of the QS. The surviving MS after post-selection is given by $|\Phi_f\rangle = \langle \psi_f|\Psi_{jt}\rangle/\sqrt{p_f}$ with the success probability $p_f = |\langle \psi_f|\Psi_{jt}\rangle|^2$. In the weak-measurement regime ($g\ll \sigma$), the final MS can be approximated to 
\begin{eqnarray}
|\Phi_f\rangle & = &\frac{1}{\sqrt{p_f}} \langle \psi_f|\exp(-ig\hat{A}\hat{P})|\psi_i\rangle|\Phi\rangle {}\nonumber\\
& \approx & \frac{1}{\sqrt{p_f}}\langle \psi_f|(1 - ig\hat{A}\hat{P})|\psi_i\rangle|\Phi\rangle{}\nonumber\\
& = & \frac{\langle \psi_f|\psi_i\rangle}{\sqrt{p_f}}(1-ig\langle \hat{A}\rangle_w\hat{P})|\Phi\rangle,
\end{eqnarray}
where the weak value is defined as
\begin{equation}
\langle \hat{A}\rangle_w = \frac{\langle \psi_f|\hat{A}|\psi_i\rangle}{\langle \psi_f|\psi_i\rangle}.
\label{eq:weak_value}
\end{equation}
If the additional AAV approximate condition 
\begin{equation}
\text{max}_{n} g\frac{|\langle \psi_f|\hat{A}^n|\psi_i\rangle|^{1/n}}{|\langle \psi_f|\psi_i\rangle|} \ll \sigma \quad \text{for}\quad n=1,2,...,
\label{eq:AAV_condition}
\end{equation}
is satisfied, then the final MS can be further simplified as
\begin{equation}
|\Phi_f\rangle \approx \exp(-ig\langle \hat{A}\rangle_w\hat{P})|\Phi\rangle.
\label{eq:final_MS}
\end{equation}

When the pre- and post-selected states approach being orthogonal, indicated by $|\langle \psi_f|\psi_i\rangle| \rightarrow 0$, the modulus of the weak value $|\langle \hat{A}\rangle_w|$ becomes significantly larger than the maximum eigenvalue of the observable $\hat{A}$. Moreover, weak value can also take complex numbers. In both of these cases, the weak value is referred to as anomalous. Both the real and imaginary parts of the large weak value can be obtained by the average value of the corresponding observables ($\hat{Q}$ or $\hat{P}$) of the MS \cite{jozsa2007complex}:
\begin{eqnarray}
\langle \hat{Q}\rangle_f &=& g\text{Re}\langle \hat{A}\rangle_w, {}\nonumber\\
\langle \hat{P}\rangle_f & =& \frac{g}{2\sigma^2}\text{Im}\langle \hat{A}\rangle_w,
\label{eq:WVA_disp1}
\end{eqnarray}
where $\langle \cdot \rangle_f$ denotes the mean value of the observable in $|\Phi_f\rangle$. This amplification form represents the initial and widely adopted form of WVA. Therefore, we refer to it as the standard WVA. Since the observables $\hat{Q}$ and $\hat{P}$ are incompatible, the measurement of either observable may not extract the complete information of a complex weak value. Researchers typically adopt a purely real or imaginary weak value in WVA. For instance, by setting the parameters of the pre- and post-selected states as $\theta_i = \pi/2 -\epsilon$, $\theta_f = -\pi/2$ and $\phi_i=\phi_f=0$ ($\theta_i = \pi/2$, $\theta_f = -\pi/2$ and $\phi_i-\phi_f = \phi$) with a small $\epsilon$ ($\phi$), a large real (imaginary) weak value $2/\epsilon$ ($-2i/\phi$) can be obtained. However, this amplification comes at the cost of a corresponding decrease in the success probability of post-selection, approximately given by $p_f \approx \epsilon^2$ and $p_f \approx \epsilon^2$ for real and imaginary weak value, respectively.

In some cases where a complex weak value is necessary, researchers have also optimized the observables of the MS to achieve the maximal amplification of the parameter $g$. Given that a complex weak value induces shifts of the MS in both $q$ and $p$, recent works \cite{knee2018seeing, pan2020weak} have utilized the Wigner function \cite{case2008wigner} to describe the final MS. This representation can be expressed as:
\begin{equation}
W_{\text{WVA}}(q,p) = p_f W_m(q-g\text{Re}\langle\hat{A}\rangle_w,p-\frac{g}{2\sigma^2}\text{Im}\langle \hat{A}\rangle_w)
\end{equation}
with $W_m(q,p) =\exp\big[-q^2/(2\sigma^2)\big]\exp(-2\sigma^2 p^2)/\pi$ denoting the Wigner function of the initial MS $|\Phi\rangle$. 

For a general observable of the MS $\hat{S}_\theta = \hat{Q}\cos\theta + \hat{P}\sin\theta$, the integration of its conjugate observable $\hat{T}_\theta = -\hat{Q}\sin\theta + \hat{P}\cos\theta$ over the distribution $W_{\text{WVA}}(q,p)$ leads to the Gaussian marginal probability distribution:
\begin{equation}
P_{\hat{S}_\theta}(s_\theta) = \frac{1}{\sqrt{2\pi\sigma_\theta^2}}\exp \Big[-\frac{[s_\theta - \langle \hat{S}_\theta\rangle_f]^2}{2\sigma_\theta^2}\Big],
\end{equation}
with the mean value $\langle \hat{S}_\theta\rangle_f$ and variance $\sigma^2_{\theta} = \sigma^2 (2\sigma^2 \cos^2\theta + \frac{1}{2\sigma^2}\sin^2\theta)$. By expressing the weak value in polar form as $\langle \hat{A}\rangle_w = |\langle \hat{A}\rangle_w|\cos\phi_{wv} + i |\langle \hat{A}\rangle_w|\sin\phi_{wv}$, the average shift of the final MS can be simplified as
\begin{equation}
\langle \hat{S}_\theta\rangle_f = g |\langle \hat{A}\rangle_w|\cos\phi_{wv} \cos\theta + \frac{g}{2\sigma^2}|\langle \hat{A}\rangle_w| \sin\phi_{wv} \sin\theta.
\end{equation}
Thus, the maximum shift of the MS $g |\langle \hat{A}\rangle_w|\sqrt{\cos^2 \phi_{wv}+\sin^2 \phi_{wv}/(4\sigma^4)}$ can be attained by satisfying the condition $2\sigma^2\tan\theta = \tan \phi_{wv}$. In Fig. \ref{fig:stdWVA} (b)-(e), the Wigner function of the initial MS $|\Phi\rangle$ and the final MS $|\Phi_f\rangle$ with various weak values are visually presented to demonstrate the optimal observables $\hat{S}_\theta$.

\subsection{2.2 Applications of weak-value amplification with photonic systems}
\label{Sec:weak measurement}

\begin{figure*}[t]
	\centering
	\includegraphics[width=0.9\textwidth]{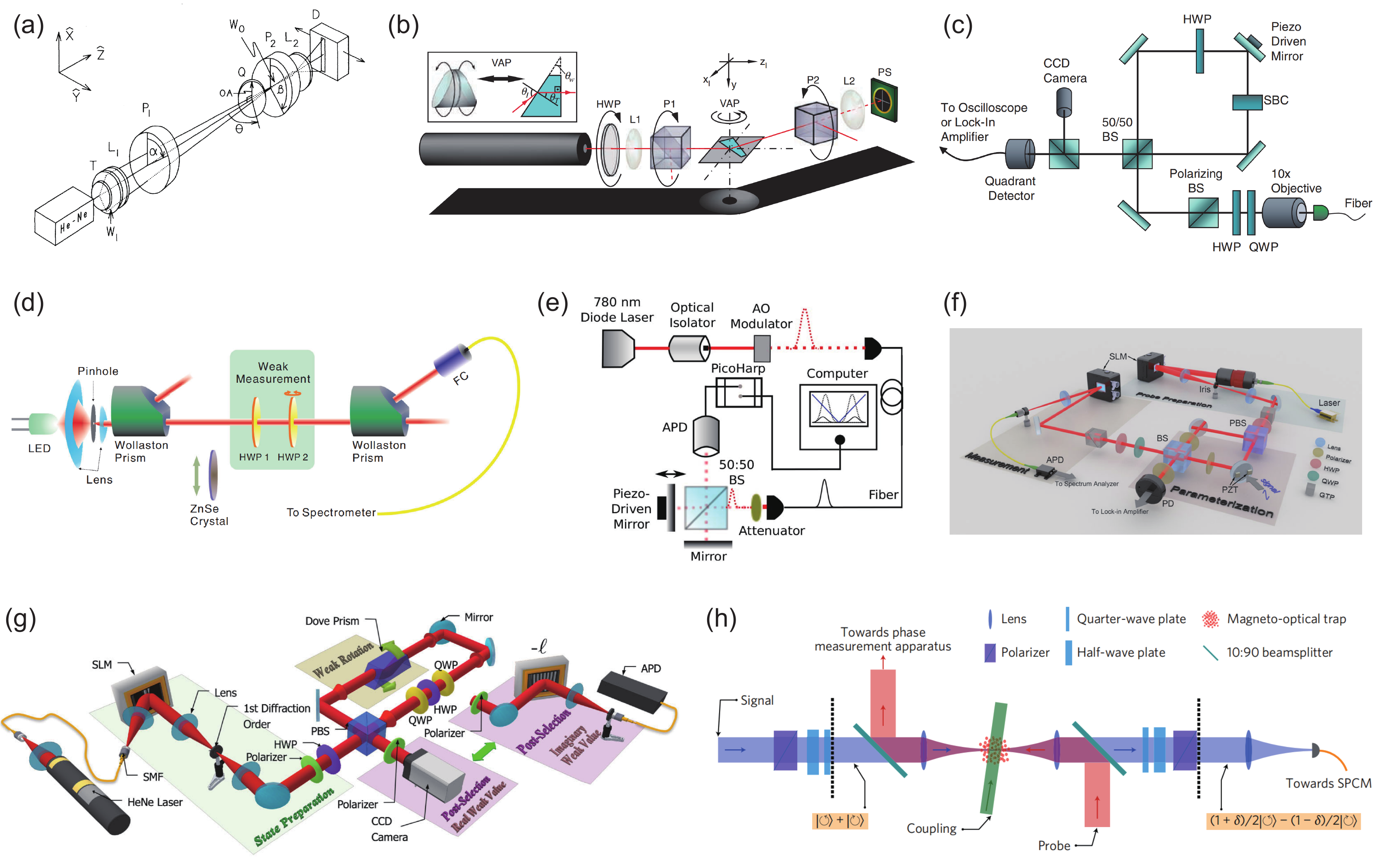}
	\caption{Applications of weak-value amplification. (a) First realization of the measurement of the weak value \cite{ritchie1991realization}. (b) The observation of the spin Hall effect of light \cite{hosten2008observation}. (c) Ultrasensitive measurement of the beam deflection \cite{dixon2009ultrasensitive}. (d) Phase estimation using a white light source \cite{brunner2010measuring}. (e) Measurement of the longitude velocities \cite{viza2013weak}. (f) achieving incompatible quantum limits on multiparameter estimation using Hermite-Gaussian states \cite{xia2023toward}. (g) Amplification of angular rotations \cite{magana2014amplification}. (h) Amplification of the non-linear effect of a single photon \cite{hallaji2017weak}.}
	\label{fig:WVAapp}
\end{figure*}

The applications of WVA primarily focus on observing ultra-small physical effects and sensing weak signals. The photonic system holds a crucial position in quantum metrology and sensing due to its good coherence, as well as its ease of manipulation and detection. Consequently, the practical applications of WVA are typically implemented by the interaction between two different degrees of freedom (DOF) of photons. The discrete variable DOF, such as polarization and path are usually referred to as the QS, while the continuous variable DOF, including the spatial, the spectrum and the temporal modes are denoted as the MS. The ultra-small coupling strength between the QS and the MS can be amplified using either the real or imaginary weak value. Fig. \ref{fig:WVAapp} illustrates the most representative applications of WVA with photonic systems.

In 1991, N. W. M. Ritchie et al. first implemented the measurement of a real weak value in an optical system \cite{ritchie1991realization} using the experimental setup depicted in Fig. \ref{fig:WVAapp} (a). The polarization and transverse spatial mode of photons are used as the QS and the MS, respectively. The transversal displacement of the light beam induced by the birefringence is amplified up to 20 times larger than its actual displacement, demonstrating the potential of weak value to amplify the small physical effects. Such interaction has been widely used to investigate the birefringence effects \cite{rhee2013amplifications, qiu2016estimation} as well as the Goos-H{\"a}nchen shift \cite{das2021quantum,zhou2021tunable}.
  
In 2008, Hosten and Kwiat utilized WVA scheme to amplify the spin Hall effect of light by four orders of magnitude, achieving a sensitivity of 1 angstrom without isolating the air disturbance and mechanical vibration \cite{hosten2008observation}. The experimental setup is illustrated in Fig. \ref{fig:WVAapp} (b). The spin, represented by the left- or right-circularly polarization, and the transverse spatial mode of photons serves as the QS and the MS, respectively. Through observing the momentum of the post-selected photons, the transversal displacement of light caused by the SHEL is amplified by the imaginary weak value. This experiment marked the first practical application of WVA in amplifying small physical effects. Subsequently, WVA has become a popular approach to study the spin Hall effect of light, as evidenced by numerous studies \cite{chen2015modified, zhou2012experimental, qin2020nanoscale, pfeifer2011weak, qin2019improved, wang2013high, gorodetski2012weak, neugebauer2018weak}. 

In 2009, Dixon et al. applied the WVA protocol to sensing the ultra-small tilt of a mirror, achieving a sensitivity of 400-frad sensitivity \cite{dixon2009ultrasensitive, howell2010interferometric} with the experimental setup depicted in Fig. \ref{fig:WVAapp} (c). The which-path information, clockwise or anticlockwise, inside the Sagnac interferometer and the transverse spatial distribution of photons are considered as the QS and the MS, respectively. A small tilt of the mirror driven by the Piezo couples the QS and the MS. The post-selection is implemented by monitoring the dark port of the interferometer. The imaginary weak value maps the ultra-small tilt angle of the mirror to the shift of the transverse beam position. This work extended the scope of the applications of WVA from studying physical effects to the sensing of practical parameters. In 2010, Starling et. al. measured the beam deflection induced by a change in optical frequency, achieving a sensitivity of $129\pm 7 \text{kHz}/\sqrt{\text{Hz}}$ with 2 mW continuous wave \cite{starling2010precision}. In 2011, a quasi-autocollimator was designed by Turner et al., which attained the sensitivity better than 10 $\text{prad}/\sqrt{\text{Hz}}$ between 10 and 200 $\text{Hz}$ \cite{turner2011picoradian}. Hogan et al. incorporated an optical lever into the WVA protocol, substantially reducing the noise floor down to 1.3 $\text{prad}/\sqrt{\text{Hz}}$ at 2.4 $\text{kHz}$ \cite{hogan2011precision}. In 2012, Egan et al. reported a temperature sensor based on WVA by implanting a prism filled with thermo-optic fluid into the interferometer, achieving the thermostat sensitivity to 0.2 mK at room temperature \cite{egan2012weak}. The above-mentioned WVA schemes mainly focus on the phase modulation of the MS during the unitary interaction. The optical loss that modulates the amplitude of the MS also carries useful information, corresponding to the nonunitary process. Liu et al. theoretically extend WVA to the nonunitary regime \cite{liu2019weak}. In 2022, Li et al. experimentally achieved the efficiency of the nonunitary WVA nearly ninefold that of unitary WVA by observing the quadratic relation between the meter shift and the norm of the weak value \cite{li2022experimental}.

In 2010, Brunner et. al. proposed a WVA protocol for measuring the longitude phase shift of photons. This approach, combining the imaginary weak value and frequency-domain measurement, was shown to outperform the CM by several orders of magnitude \cite{brunner2010measuring}. In 2013, Xu et al. achieved high-precision phase estimation using a white light source \cite{xu2013phase}, as illustrated in Fig. \ref{fig:WVAapp} (d). The polarization and frequency spectrum of photons are denoted as the QS and the MS, respectively. The wide spectrum distribution of the white light source significantly contributes to the amplification effect of the longitude phase. The experimental results also demonstrate the robustness of phase estimation based on imaginary WVA against chromatic dispersion. This configuration of the QS and the MS is commonly employed in measuring relative time delay caused by temperature shifts \cite{salazar2015enhancement}, rotatory dispersion effects \cite{zhou2020measuring}, glucose concentration \cite{li2016application} and other effects \cite{fang2016ultra, salazar2014observation, salazar2014measurement, li2021quantification, zanetto2021electrical, huang2021weak, wang2021realization, zhong2021simultaneous, liu2021high}. 

In 2013, Viza et al. demonstrated an imaginary WVA scheme capable of measuring the Doppler frequency shift produced by a moving mirror in the time domain with the experimental setup shown in Fig. \ref{fig:WVAapp} (e) \cite{viza2013weak}. The which-path information in a Michelson Interferometer and the non-Fourier limited Gaussian temporal pulse of photons generated by the acousto-optic modulator are referred to as the QS and the MS, respectively. For the MS, the temporal length of the pulse can be much larger than the coherence length of the laser. The reported sensitivity of the velocity measurement is down to 400 $\text{fs}/\text{s}$. Recent studies have improved the sensitivity and SNR of the velocity measurement using the Vernier-effect principal \cite{huang2021improving}. The time-domain measurement of the imaginary WVA can be also applied to measure ultra-small parameters such as group delay \cite{mirhosseini2016weak} and angular velocity \cite{fang2021weak}, and demonstrate interesting physical effects like fast-light effect \cite{mirhosseini2016weak}. The applications related to the Fig. \ref{fig:WVAapp} (d) and (e) indicate that the incoherent MS can contribute to the amplification of the coupling strength when combined with the imaginary weak value.

Apart from the conventional Gaussian-shaped meter state, the high-order Hermite-Gaussian and Laguerre-Gaussian modes provide WVA with significant potential for measuring complex signals. In Fig. \ref{fig:WVAapp} (f), Xia et al. proposed a multi-parameter WVA scheme using $n$-order Hermite-Gaussian states as the MS \cite{xia2020high, xia2023toward}. This approach improves the precision by a factor of $2n+1$ compared to the Gaussian MS. Moreover, it enables simultaneous estimation of spatial displacement and angular tilt of light, achieving the precisions up to 1.45 nm and 4.08 nrad, respectively. In Fig. \ref{fig:WVAapp} (g), Maga$\tilde{\text{n}}$a-Loaiza et al. demonstrated WVA in the azimuthal DOF \cite{magana2014amplification}, estimating the angular rotation by utilizing the polarization DOF and the spatial mode with orbital angular momentum of photons as the QS and the MS, respectively. The rotation of a Dove prism inside the Sagnac interferometer that implements the spin-orbit coupling is amplified by about two orders of magnitude.

In most applications of WVA as listed above, the coupling between the QS and the MS is typically achieved through the interaction between different DOFs of single photons. This scenario of WVA can be explained using classical electromagnetic theory, which has raised a debate regarding the quantum nature of weak values. To address this argument, researchers have attempted to measure weak values without resort to classical interactions. For example, the observation of anomalous weak values has been demonstrated in massive-particle systems \cite{sponar2015weak}. In 2005, Pryde et al. successfully measured weak values of photons based on nonclassical interaction, e.g., Hong-Ou-Mandle interference between photons. In 2013, Feizpour et. al. proposed the theoretical WVA protocol to amplify the optical non-linearities at the single-photon level \cite{feizpour2011amplifying}. In 2017, Hallaji et al. experimentally achieved an eightfold amplification of the non-linear phase shift caused by a single photon \cite{hallaji2017weak} with the experimental setup depicted in Fig. \ref{fig:WVAapp} (h). The amplification of the nonlinear interaction strength has also inspired practical Heisenberg-scaling metrology without the need for quantum resources \cite{zhang2015precision, chen2018achieving, chen2018heisenberg}. In order to clarify the role of statistical averaging in generating anomalous weak values, Rebufello et al. designed a single-click experiment to obtain the anomalous weak value with the amplified outcomes largely surpassing the uncertainty of the MS \cite{rebufello2021anomalous}. This finding enriches the meaning of weak values beyond a statistical concept of conditional expectation value, thus paving the way for the robust WVA applications.

\subsection{2.3 Generalization of weak-value amplification}

In the standard WVA, whether the shifts of post-selected MS are proportional to weak values relies on the validity of two crucial approximate conditions, namely, the weak-measurement approximation ($g\ll\sigma$) and the AAV approximation condition presented in Eq. \eqref{eq:AAV_condition}. The former provides the possibility of generating anomalous weak values that surpass the eigenvalue spectrum of the observable. As a contrast, the latter determines that the amplified shifts in the final MS cannot be arbitrarily large for a fixed coupling strength. The generalization of the standard WVA relies on rigorously deriving the shifts of the post-selected MS without approximations. In 2011, Koike et. al. conducted a study on the practical limits of WVA by considering all orders of the coupling strength $g$. They derived the optimal overlap of the pre- and post-selected states to achieve the maximum shifts of the post-selected MS on the Sagnac interferometer \cite{koike2011limits}. Zhu et al. derived the rigorous expressions for the shifts of the post-selected MS in WVA without restricting the coupling strength. The maximum shifts of MS in WVA are also obtained for a qubit system \cite{zhu2011quantum}. Wu and pang et al. focused on the WVA protocol with asymptotically or exactly orthogonal pre- and post-selected states \cite{pang2012weak, wu2011weak}. A general rigorous framework for the weak measurement beyond the AAV approximation condition was established. Depending on the values of the coupling strength $g$, the standard derivation of the MS $\sigma$ and the weak value $\langle \hat{A}\rangle_w$, the actual shifts of the post-selected MS can be classified into three regions: (i) strong-measurement region ($g \gg \sigma$); (ii) WVA region ($g<g|\langle \hat{A}\rangle_w|\ll \sigma$); (iii) inverse WVA region ($g\ll \sigma \ll g|\langle \hat{A}\rangle_w|$). In the following, we adopt the derivations of Wu et al. \cite{wu2011weak} to clarify the generalization of the WVA scheme.

With a general pre-selected state $\rho_s$ and a measurement operator $\hat{\Pi}_f$ implementing the post-selection, Wu et al. defined the high-order weak value of the observable $\hat{A}$ as
\begin{equation}
\langle \hat{A}\rangle^{m,l}_w = \frac{\text{Tr}(\hat{\Pi}_f \hat{A}^m \rho_s \hat{A}^l)}{\text{Tr}(\hat{\Pi}_f\rho_s)},
\end{equation}
By taking up to the second-order approximation about the coupling strength $g$, the shifts of the final MS can be expressed with the high-order weak values as follows:
\begin{eqnarray}
\langle \hat{Q}\rangle_f & \approx & \frac{g\text{Re} \langle \hat{A}\rangle_w + g\text{Im} \langle \hat{A}\rangle_w \langle \{\hat{Q},\hat{P}\}\rangle }{1+g^2 \text{Var}(\hat{P})(\langle \hat{A}\rangle_w^{1,1}-\text{Re}\langle \hat{A}^2\rangle_w)} = \frac{4g\text{Re}\langle \hat{A}\rangle_w\sigma^2}{4\sigma^2 + g^2(|\langle \hat{A}\rangle_w|^2 - 1)}{}\nonumber\\
\langle \hat{P}\rangle_f &\approx & \frac{2g\text{Im} \langle \hat{A}\rangle_w \text{Var}(\hat{P}) }{1+g^2 \text{Var}(\hat{P})(\langle \hat{A}\rangle_w^{1,1}-\text{Re}\langle \hat{A}^2\rangle_w)} = \frac{2g\text{Im}\langle \hat{A}\rangle_w}{4\sigma^2 + g^2(|\langle \hat{A}\rangle_w|^2-1)}.
\label{eq:wva_disp_exact}
\end{eqnarray}
Here, the simplified results are obtained by considering the observable $\hat{A} = \hat{\sigma}_z = |0\rangle\langle 0| - |1\rangle\langle 1|$ and the Gaussian MS in Eq. \eqref{eq:initial_MS}. When the coupling strength $g$ and weak value $\langle \hat{A}\rangle_w$ satisfy both the weak-measurement and AAV approximation conditions, Eq. \eqref{eq:wva_disp_exact} degrades to Eq. \eqref{eq:WVA_disp1}, corresponding to the standard WVA. However, increasing the modular of weak value continuously can lead to a violation of the AAV approximation condition. This results in an initial increase followed by a rapid decrease in the shift of the post-selected MS. During this process, the measurement scheme transitions from the WVA region to the inverse WVA region. The transition points correspond to the maximum values of $\langle \hat{Q}\rangle_f$ and $\langle \hat{P}\rangle_f$ in the real and imaginary WVA schemes, respectively. According to Eq. \eqref{eq:wva_disp_exact}, these maximum values are derived as $\max \langle \hat{Q}\rangle_f = \sigma$ and $\max \langle \hat{P}\rangle_f = 1/(2\sigma)$, which aligns well with the rigorous derivations by Zhu et. al. \cite{zhu2011quantum}. When $|\langle \hat{A}\rangle_w|$ increases such that $g|\langle \hat{A}\rangle_w|\gg \sigma$, Eq. \eqref{eq:wva_disp_exact} can be approximated as $\langle \hat{Q}\rangle_f \approx 4\sigma^2 / (g \text{Re}\langle \hat{A}\rangle_w)$ and $\langle \hat{P}\rangle_f \approx 2/(g\text{Im}\langle \hat{A}\rangle_w)$ for the real and imaginary weak values, respectively. In such cases, the shifts $\langle \hat{Q}\rangle_f$ and $\langle \hat{P}\rangle_f$ are inversely proportional to the weak value, and the measurement protocol operates in the inverse WVA region. In this region, the weak value and the associated parameters (e.g., $\theta_i$, $\theta_f$) in the QS can be effectively amplified by a factor of $1/g$, which will be discussed further in section 6.

\begin{figure*}[t]
	\centering
	\includegraphics[width=0.9\textwidth]{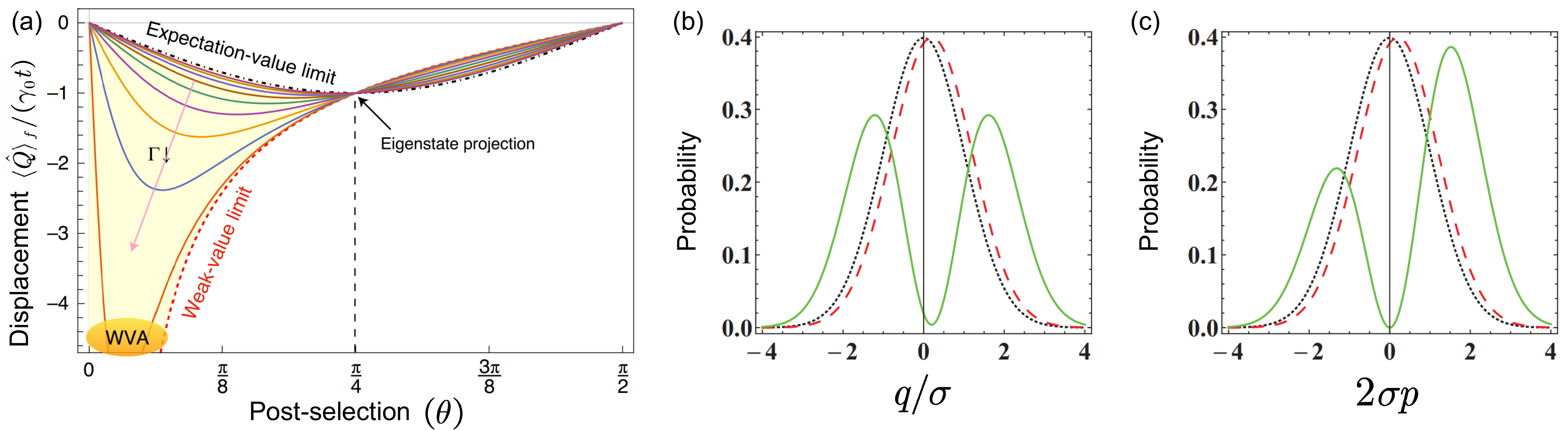}
	\caption{(a) The normalized shifts of the final meter state $\langle \hat{Q}\rangle_f/(\gamma_0 t)$ are plotted with different post-selection angle $\theta$ during the weak-to-strong measurement transition in a trapped-ion system. In (b) and (c), the probability density of Gaussian initial meter states (the dotted black lines), the post-selected meter states in standard (the dashed red lines) and orthogonal (the solid green lines) WVA are compared in $q$ and $p$ representations, respectively. In (b) and (c), both the weak value $\langle \hat{A}\rangle_w$ and the orthogonal weak value $\langle \hat{A}\rangle_{ow}^{1,0}$ are set to $0.2+0.1i$.}
	\label{fig:general_WVA}
\end{figure*}

The weak-to-strong measurement transition of the pre- and post-selected QS was experimentally examined in a trap-ion system in 2020 \cite{pan2020weak}. We adopt their experimental parameters to demonstrate the actual shifts of the final MS within all three regions and two critical points between them. In their work, the QS utilized the internal electronic state of a single trapped $^{40}\text{Ca}^+$ ion, while the axial vibrational motion of the ion serves as the MS. The pre-selected state of the QS, denoted as $|\psi_i\rangle = |1\rangle$, interacts with the Gaussian MS $|\Phi\rangle$ under the Hamiltonian $\hat{H} = \gamma_0 \hat{\sigma}_x \otimes \hat{P}$, where $\hat{\sigma}_x = |1\rangle\langle 0| + |0\rangle\langle 1|$ and $\hat{P}$ represent the observables of the QS and the MS, respectively. The interaction time $t$ can be adjusted to control the coupling strength $g = \gamma_0 t$. The joint state is then post-selected by $|\psi_f\rangle = \cos\theta |0\rangle - \sin\theta |1\rangle$, resulting in a real weak value $\langle \hat{\sigma}_x\rangle_w = -\cot \theta$. Consequently, the shift of the post-selected MS in $q$ space is given by
\begin{equation}
\langle \hat{Q}\rangle_f = -\frac{\gamma_0 t\sin 2\theta}{1-\cos(2\theta)e^{-\Gamma^2/2}},
\end{equation}
where $\Gamma = \gamma_0 t/\sigma$ characterizes the relative coupling strength. The average shift $\langle \hat{Q}\rangle_f$ during the strong-to-weak measurement transition are illustrated in Fig. \ref{fig:general_WVA} (a) with different coupling strength $\Gamma$ and the parameter $\theta$ in the post-selection.  

When the coupling strength is strong ($\Gamma \gg 1$), the shift of the final MS can be approximated as $\langle \hat{Q}\rangle_f|_{\Gamma \rightarrow \inf} = -\gamma_0 t \sin2\theta = \gamma_0 t \langle \psi_f| \hat{\sigma}_x|\psi_f\rangle$, which is directly proportional to the expectation value of the observable $\hat{\sigma}_x$ in the post-selected state $|\psi_f\rangle$. In this strong measurement region, the ratio of $\langle \hat{Q}\rangle_f$ to $\gamma_0 t$ remains within the eigenvalue spectrum of the observable. As the coupling strength weakens to satisfy the AAV condition ($\Gamma |\langle \hat{\sigma}_x\rangle_w| \ll 1$), the ratio $\langle \hat{Q}\rangle_f/(\gamma_0 t)_{|\Gamma \rightarrow 0} \approx -\cot\theta \approx \langle \hat{\sigma}_x\rangle_w$ approaches the weak value. Furthermore, as the parameter $\theta$ decreases, the AAV approximate condition is violated due to a large weak value. The shift $\langle \hat{Q}\rangle_f$ gradually deviates from the standard WVA. This trend suggests an extreme value of $\langle \hat{Q}\rangle_f$, which can be achieved by $\theta = \text{arccos}(e^{-\Gamma^2/2})/2$. After reaching this extreme value, the shift of the MS enters the inverse WVA region, where $\langle \hat{Q}\rangle_f \approx -4\sigma^2 \cot\theta/g$.

Besides the previously mentioned regions, there exists a special case where the pre- and post-selected states are strictly orthogonal, i.e., $\langle \psi_f|\psi_i\rangle=0$. In this scenario, Wu et.al. introduced the concept of orthogonal weak value to describe the shifts of the post-selected MS \cite{wu2011weak}. With a general pre-selected state $\rho_i$ and post-selected operator $\hat{\Pi}_f$ satisfying $\text{Tr}(\rho_i \hat{\Pi}_f)=0$, the orthogonal weak value of the observable $\hat{A}$ is defined as 
\begin{equation}
\langle \hat{A}\rangle_{ow}^{m,l} = \frac{\text{Tr}(\hat{\Pi}_f \hat{A}^{m+1}\rho_i\hat{A}^{l+1})}{(m+1)(l+1)\text{Tr}(\hat{\Pi}_f \hat{A}\rho_i\hat{A})}.
\end{equation}
When considering a Gaussian initial MS $|\Phi\rangle$, the shifts of the post-selected MS in the orthogonal weak measurement scenario can be expressed similarly to the standard WVA, given by $\langle \hat{Q}\rangle_f = g \text{Re}\langle \hat{A}\rangle_{ow}^{1,0}$ and $\langle \hat{P}\rangle_f = 3g \text{Im}\langle \hat{A}\rangle_{ow}^{1,0}/(2\sigma^2)$. The probability distribution of the post-selected MS in both the standard WVA and the orthogonal weak measurement can be compared. As shown in Fig. \ref{fig:general_WVA} (b) and (c), the average shifts $\langle \hat{Q}\rangle_f$ and $\langle \hat{P}\rangle_f$ in orthogonal weak measurement result from the displacement and the relative intensity variation of bimodal-shape distribution, respectively.

Apart from the average shift of the post-selected MS, Parks et.al. derived the variance of an arbitrary observable $\hat{M}$ under the post-selected MS $|\Phi_f\rangle$ with a complex weak value. They found that the variance of the observable $\hat{P}$ and $\hat{Q}$ are only related to the imaginary part of weak value and third central moment of $\hat{P}$ relative to the initial MS \cite{parks2011variance}. The exact formulas that include all-order effects of the coupling strength with a typical observables $\hat{A}$ (satisfying $\hat{A}^2=\hat{I}$) are provided to evaluate the shifts and the probability distributions of post-selected MS \cite{PhysRevA.85.012113}.  Additionally, Lorenzo et. al. derived the characteristic function for the moments of the post-selected MS to fully describe the statistics of WVA protocol \cite{di2012full}. For a general derivation of the Von Neumann measurement in the framework of pre- and post-selection, we suggest to refer to the review paper Ref. \cite{kofman2012nonperturbative} for further reading.

Though the Gaussian superposition state $|\Phi\rangle$ is most prevalent in both the theoretical investigations and experimental applications, the WVA scheme can also be implemented with more general types of MS. In 2004, Johansen has provided the derivation of weak value with the arbitrary MS including the mixed states \cite{johansen2004weak}. Moreover, the wavefunction of the MS can be optimized to acquire the maximal amplification factor in WVA for a given weak value, which removes the maximal shifts restricted by the Gaussian MS \cite{susa2012optimal}. Specifically, the theoretical schemes were developed to realize WVA with the MS such as a qubit \cite{wu2009weak}, orbital-angular-momentum states \cite{puentes2012weak}, Hermite-Gaussian states, Laguerre-Gaussian states \cite{turek2015post}, thermal states, mixed states \cite{cho2010weak} and etc \cite{luo2020anomalous}. The concept of modular values are proposed to describe the coupling between the qubit MS and the QS with the arbitrary strength \cite{kedem2010modular}. In order to adapt the requirements of WVA in more general cases, e.g., in the presence of the decoherence, Shikano et. al. introduced a $\hat{W}$ operator and the associated quantum operations on it to formally describe the weak value \cite{shikano2009weak}. Recent work has also shown that the weak value can be operationally formulated as the response of the pre- and post-selected QS without resort to the MS \cite{ogawa2020operational}.

\section{3. Precision analysis on the weak-value amplification}

As WVA scheme plays an increasingly crucial role in practical precision metrology, it is imperative to develop generic approaches to evaluate its precision. This research not only improves our understanding of the potential advantages of WVA but also provides guidance for further enhancing its performance. In this section, we first review the quantum metrology theory for single-parameter estimation and introduce the Fisher information (FI) metric as a means to determine the ultimate precision of general measurement protocols. Moreover, we compare the widely used signal-to-noise ratio (SNR) metric in the field of sensing with FI, clarifying their consistency and differences in evaluating measurement protocols. Subsequently, we apply the FI metric to analyze the ultimate precision of WVA and discuss the attainability of its optimal precision.

\subsection{3.1 Quantum metrology theory}

\begin{figure*}[t]
	\centering
	\includegraphics[width=0.95\textwidth]{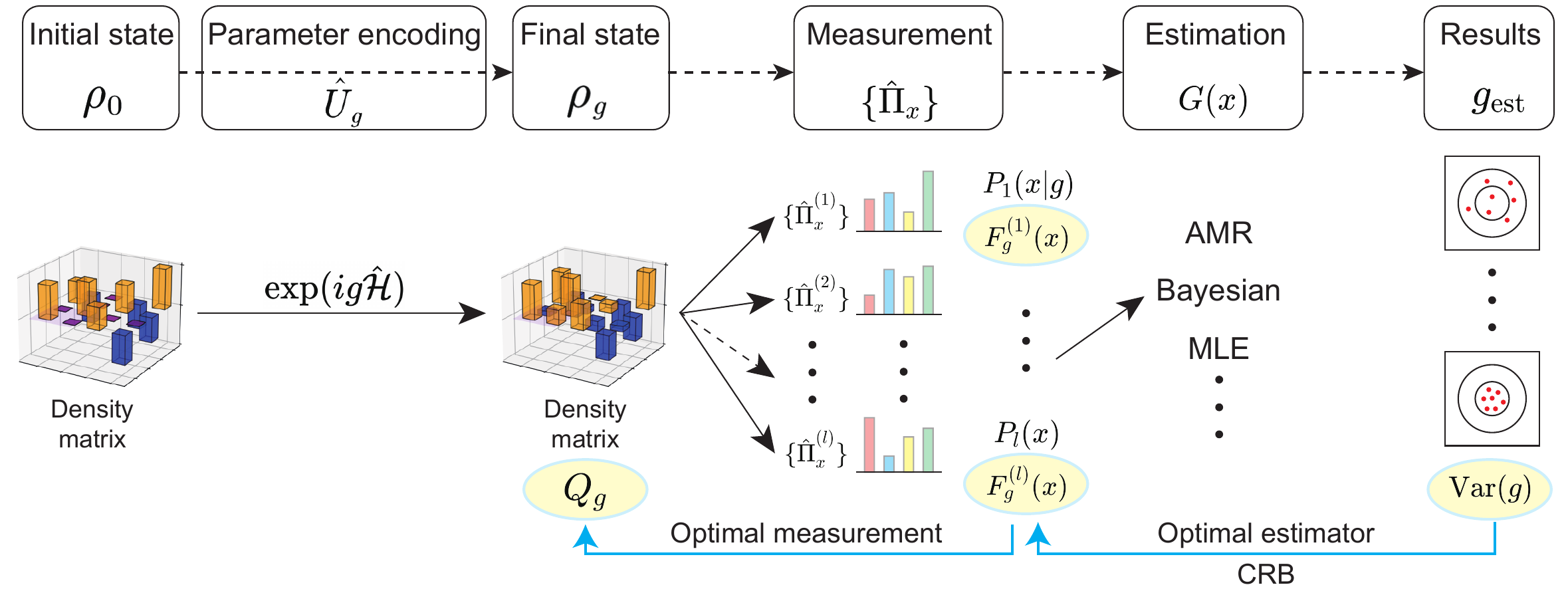}
	\caption{The basic procedure of quantum metrology for single-parameter estimation. Some related abbreviations: averaging the measurement results (AMR), maximum likelihood estimation (MLE), Cram\'er-Rao bound (CRB), standard quantum limit (SQL), Heisenberg limit (HL).}
\label{fig:metrology}
\end{figure*}

In quantum metrology, the general procedure to estimate an unknown parameter $g$ is presented in Fig. \ref{fig:metrology} \cite{giovannetti2011advances, degen2017quantum, polino2020photonic}. Initially, a quantum state $\rho_0$ is prepared, which evolves under the parameter-encoding process $\hat{U}_g$ to obtain the final state $\rho_g = \hat{U}_g\rho_0\hat{U}_g^\dagger$. Subsequently, a general quantum measurement, known as a positive-operator-valued measure (POVM) $\{\hat{\Pi}_x\}$ is performed on the final state to transfer the information about the parameter $g$ from $\rho_g$ to the probability distribution $P(x|g)$. The conditional probability for each measurement result $x$ is determined by Born's rule: $P(x|g) = \text{Tr}(\rho_g\hat{\Pi}_x)$. After repeating the measurement process $\nu$ times, a sequence of measurement results $\vec{x}=(x_1,...,x_\nu)$ is obtained. Finally, the parameter is estimated using an appropriate estimator $g_{\text{est}} = G(\vec{x})$.

The precision of parameter estimation is typically assessed by the variance of the estimated parameter, denoted as $\text{Var}(\cdot)$. The ultimate precision of unbiased estimation of the parameter $g$ from the sequence $\vec{x}$ is determined by the Cram\'er-Rao bound (CRB):
\begin{equation}
\text{Var} (g) \ge \frac{1}{\nu F_g}.
\end{equation}
Here, the Fisher information (FI), defined as:
\begin{equation}
F_g = \int \frac{1}{P(x|g)} \Big[ \frac{\partial P(x|g)}{\partial g}\Big]^2 dx,
\label{eq:FI_def}
\end{equation}
quantifies the sensitivity of the probability distribution $P(x|g)$ to the variations in the parameter $g$ \cite{fisher1922mathematical}. A larger FI indicates a potential for achieving better estimation precision. The CRB can be asymptotically attained with a sufficiently large number of measurements ($\nu$) and an efficient estimator, such as the maximum likelihood estimation (MLE) \cite{lehmann2006theory}, Bayesian estimation \cite{pezze2014quantum, trees2007bayesian, le2012asymptotic}, etc. For a given final state $\rho_g$, the probability distribution $P_l(x|g)$ varies as the POVM $\{\hat{\Pi}^{(l)}_x\}$ changes, resulting in different FI $F_g^{(l)}$. By maximizing the FI over all possible POVM, we can obtain the quantum Fisher information (QFI): $Q_g = \max_{\{\hat{\Pi}_x^{(l)}\}} F_g$ \cite{lee2002quantum}. The QFI quantifies the sensitivity of the parameter-encoded quantum state $\rho_g$ to the parameter $g$. When both the quantum measurement is optimal and the estimator is efficient, the precision of estimating the parameter $g$ can achieve the quantum CRB $\Delta^2g \ge 1/(\nu Q_g)$. 

To rigorously determine the QFI $Q_g$ of a general state $\rho_g$, a super operator called the symmetric logarithmic derivative (SLD) $\mathcal{R}_{\rho_g}^{-1}(\partial_g \rho_g)$ is introduced with the definition $\partial_g \rho_g = [\rho_g\mathcal{R}_{\rho_g}^{-1}(\partial_g \rho_g) + \mathcal{R}_{\rho_g}^{-1}(\partial_g \rho_g) \rho_g]/2$. By expressing the SLD $\mathcal{R}_{\rho_g}^{-1}(O)$ in the diagonal representation of $\rho_g =\sum_{j}\lambda_j |j\rangle\langle j| $ as
\begin{equation}
\mathcal{R}_{\rho_g}^{-1}(O) = \sum_{j,k:\lambda_j+\lambda_k \neq 0}\frac{2O_{jk}|j\rangle\langle k|}{\lambda_j + \lambda_k},
\end{equation}
we can obtain the QFI as \cite{helstrom1969quantum, holevo2011probabilistic, lee2002quantum}
\begin{equation}
Q_g = \text{Tr}[ \mathcal{R}_{\rho_g}^{-1}(\partial_g \rho_g)\rho_g \mathcal{R}_{\rho_g}^{-1}(\partial_g \rho_g)].
\label{eq:QFI}
\end{equation}
The definition of the QFI provides it with two fundamental properties: convexity and additivity: (i) For a mixed state $\rho_g = \sum_{j}c_j \rho_g^{(j)}$, we have $Q_g\le \sum_{j} c_j Q_g^{(j)}$. (ii) For a direct product state $\rho_g = \otimes _j \rho_g^{(j)}$, we have $Q_g = \sum_j Q_g^{(j)}$, where $Q_g^{(j)}$ represents the QFI of $\rho_g^{(j)}$. These properties allows for the convenient derivation of upper bounds for the QFI in compound quantum systems. When $\rho_g = |\psi_g\rangle\langle \psi_g|$ is a pure state, the expression for the QFI simplifies to:
\begin{equation}
Q_g = 4\Big[ \frac{d\langle \psi_g|}{dg}\frac{d|\psi_g\rangle}{dg}  - \Big|\frac{d\langle \psi_g|}{dg}|\psi_g\rangle \Big|^2 \Big].
\label{eq:QFI_pure}
\end{equation}
If we consider the parameter-encoding process as a unitary evolution $\hat{U}=\exp(-ig\hat{\mathcal{H}})$ with the generator $\hat{\mathcal{H}}$, the QFI depends solely on the pure initial state $\rho_0 = |\psi_0\rangle\langle \psi_0|$:
\begin{equation}
Q_g = 4[\langle \psi_0|\hat{\mathcal{H}}^2|\psi_0\rangle - \langle \psi_0|\hat{\mathcal{H}}|\psi_0\rangle^2] = 4\text{Var}_{|\psi_0\rangle}(\hat{\mathcal{H}}),
\end{equation}
Here, the quantum CRB coincides with the generalized uncertainty relation $\text{Var}(g)\text{Var}_{|\psi_0\rangle}(\hat{\mathcal{H}})\ge 1/4$. Both the QFI and the uncertainty relation indicate that the optimal precision of $g$ is achieved with the maximum variance $\text{Var}_{|\psi_0\rangle}(\hat{\mathcal{H}})$ \cite{braunstein1996generalized}.

\begin{figure*}[t]
	\centering
	\includegraphics[width=0.95\textwidth]{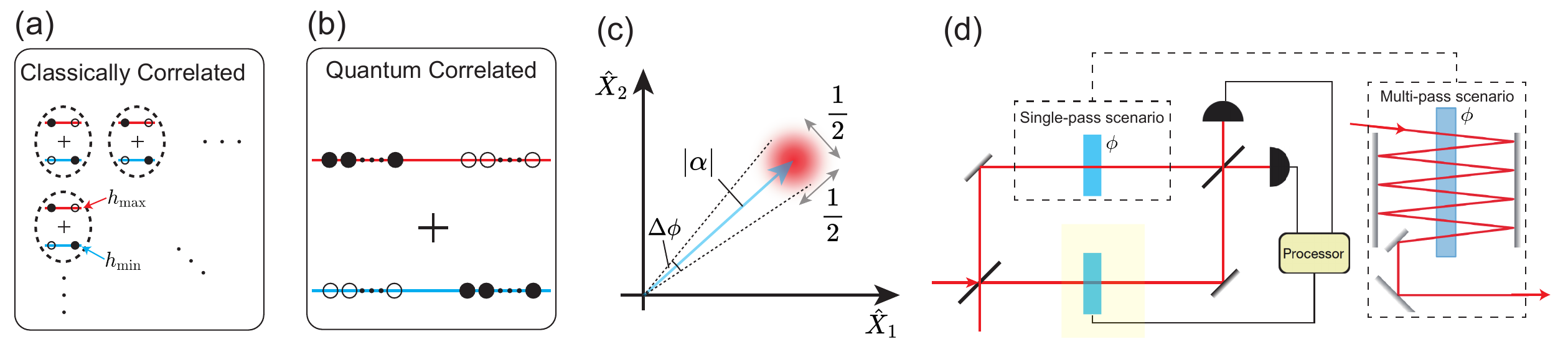}
	\caption{(a) $N$ classically correlated states are employed to measure the parameter $g$ with the precision attaining the standard quantum limit (SQL). (b) A maximally entangled state is employed to measure the parameter $g$ with the precision achieving Heisenberg limit. (c) The phase space distribution of the coherent state. (d) A Mach-Zender interferometer that measures an unknown phase $\phi$ with single-pass and multi-pass scenarios \cite{higgins2007entanglement}.}
\label{fig:Q_limit}
\end{figure*}

The linear parameter-encoding process $\hat{U}_g$ introduces fundamental bounds on measurement precision when optimizing the initial state $|\psi_0\rangle$. Fig. \ref{fig:Q_limit} (a) illustrates $N$ classically correlated states $|\psi_0^{(j)}\rangle$ ($j=1,2,...,N$) composing the total product state $|\Psi_0\rangle_{\text{tot}}^{(c)} = \otimes _{j=1}^N |\psi_0^{(j)}\rangle$. By considering the maximal ($h_{\text{max}}$) and minimal ($h_{\text{min}}$) eigenvalues of the generator $\hat{\mathcal{H}}$, the optimal state $|\psi_0^{(j)}\rangle = (|h_\text{min}\rangle + |h_\text{max}\rangle)/\sqrt{2}$ yields a maximum QFI of $Q_g^{(j)} = (h_\text{max}-h_\text{min})^2$. When $N$ copies of the optimal states are used, the total QFI is maximized as follows:
\begin{equation}
Q_{\text{tot}}^{(c)} = 4\text{Var}_{|\Psi_0\rangle_{\text{tot}}^{(c)}}(\hat{\mathcal{H}}^{\otimes N}) = N(h_{\text{max}} - h_{\text{min}})^2.
\end{equation}
Consequently, the precision of $g$ ultimately scales as $\text{Var} (g) \propto 1/N$ \cite{giovannetti2006quantum, pezze2009entanglement}, similar to the central limit theorem, which is referred to as the standard quantum limit (SQL). Then, we consider a quantum-correlated system denoted by the total state $|\Psi_0\rangle_{\text{tot}}^{(q)}$. By globally maximizing the total QFI, we obtain:
\begin{equation}
Q_{\text{tot}}^{(q)} = \text{max} \text{Var}_{|\Psi_0\rangle_{\text{tot}}^{(q)}}(\hat{\mathcal{H}}) = N^2(h_{\text{max}}-h_{\text{min}})^2.
\end{equation}
The corresponding optimal state is the maximally entangled state $|\Psi_0\rangle_{\text{tot}}^{(q)} = (|h_\text{max}\rangle^N + |h_\text{min}\rangle^N )/\sqrt{2}$, shown in Fig. \ref{fig:Q_limit} (b). In this case, the precision of $g$ scales as $\text{Var}(g)\propto 1/N^2$ \cite{berry2000optimal, gorecki2020pi, buvzek1999optimal}. Compared to the SQL, the maximally entangled state enhances the precision limit by a factor of $N$, known as the Heisenberg Limit (HL).

In Fig. \ref{fig:Q_limit} (d), we present the measurement of the relative phase $\phi$ between two paths (labelled as 1 and 2) in a Mach-Zehnder interferometer. The generator for the parameter $\phi$ is the photon number operator associated with the relative path, denoted as $\hat{\mathcal{H}} = \hat{n}_1$. As discussed earlier, the SQL and HL can be achieved by using $N$ independent single photons in the state $|\psi_N\rangle = 2^{-N/2}(|1\rangle_1|0\rangle_2 + |0\rangle_1|1\rangle_2)^{\otimes N}$ and the `N00N' state $|\psi_{N00N}\rangle = (|N\rangle_1|0\rangle_2 + |0\rangle_1|N\rangle_2)/\sqrt{2}$, respectively. Another commonly used classical photonic state is the coherent state $|\alpha\rangle$, whose phase-space distribution is shown in Fig. \ref{fig:Q_limit} (c). When estimating the relative phase $\phi$ using the coherent state, the precision is ultimately limited by the variance of the photon number operator $\hat{n}$ in $|\alpha\rangle$: $\text{Var}(\phi) \propto 1/\text{Var}(\hat{n})_{|\alpha\rangle} = 1/\bar{n}$, where $\bar{n} = |\langle \alpha|\alpha\rangle|^2$ is the mean photon number. This precision limit is referred to as the shot noise limit (SNL) \cite{braunstein1992quantum}. To reduce the photon number fluctuations, the phase-space distribution can be squeezed along a specific direction. The squeezing operator $S_2(r) = \exp(r^* \hat{a}_1\hat{a}_2 - r\hat{a}_1^\dagger \hat{a}_2^\dagger)$ acting on the two-mode vacuum state leads to the two-mode squeezed vacuum states $|r\rangle_2 = S_2(r) |0\rangle_1|0\rangle_2$. The average and variance of the photon number operator in $|r\rangle_2$ are given by $\bar{n} = 2\sinh^2 |r|$ and $\text{Var}_{|r\rangle_2}(\hat{n}_1) = 2(\bar{n}^2 + \bar{n})$, respectively. Consequently, the precision of $\phi$ can be improved to $\text{Var}(\phi) = 1/[8(\bar{n}^2 + \bar{n})]$, attaining Heisenberg-scaling precision with large $\bar{n}$ \cite{yurke19862}. However, the preparation of quantum-correlated states (e.g., N00N states and squeezed states) with large photon numbers remains a significant experimental challenge, hindering the practical implementation of quantum metrology enhanced by quantum resources. To circumvent these technical problems, Higgins et al. proposed an entanglement-free Heisenberg-limited phase-estimation protocol by replacing the single-pass phase shift with the multiple-pass scenario, as shown in Fig. \ref{fig:Q_limit} (d). 

The signal-to-noise ratio (SNR) is also a commonly used metric for evaluating the performance of metrological protocols, which depends significantly on the specific measurement schemes \cite{barnett2003ultimate}. For instance, we consider the conventional scheme (CM) to measure the transversal displacement of a Gaussian beam $|\Phi\rangle$ due to the unitary evolution $\hat{U}_g = \exp(-ig\hat{P})$. The final state $|\Phi_g\rangle$ is measured with the POVM $\{\hat{\Pi}_x\}$, which yields the measurement result $x$ subjecting to the probability distribution $P(x|g) = \langle \Phi_g|\hat{\Pi}_x|\Phi_g\rangle$. The terms "signal" and "noise" correspond to the average shift and the standard deviation of $x$, respectively. Therefore, the SNR of total $\nu$ measurements is given by
\begin{equation}
\text{SNR} =\frac{\sqrt{\nu}|\langle x\rangle - x_0|}{\sqrt{\langle x^2\rangle - \langle x\rangle^2}},
\label{eq:SNR_def}
\end{equation}
where $\langle x\rangle = \int P(x|g) x dx$, and $x_0 = \langle \Phi|\hat{\Pi}_x|\Phi\rangle$ is the average of $x$ for the initial state. The SNR metric naturally defines both the quantum measurement $\{\hat{\Pi}_x\}$ and the estimator of averaging the measurement results (AMR) \cite{pang2015improving}. If the AMR estimator efficiently saturates the CRB, the SNR metric is equivalent to the FI in quantifying the metrological information of the parameter in a probability distribution $P(x|g)$. In this scenario, the MLE is equivalent to the AMR estimator with a Gaussian MS $|\Phi\rangle$. Specifically, the SNR can be related to the FI $F_g$ about $g$ through $\text{SNR} = \sqrt{\nu}g/\sigma = g\sqrt{\nu F_g}$. Furthermore, if $\hat{\Pi}_x$ is also the optimal quantum measurement to attain the QFI, the SNR metric captures the whole QFI of the final state $|\Phi_g\rangle$.

\subsection{3.2 Metrological Limits of Weak Value Amplification}
\begin{figure*}[t]
	\centering
	\includegraphics[width=0.95\textwidth]{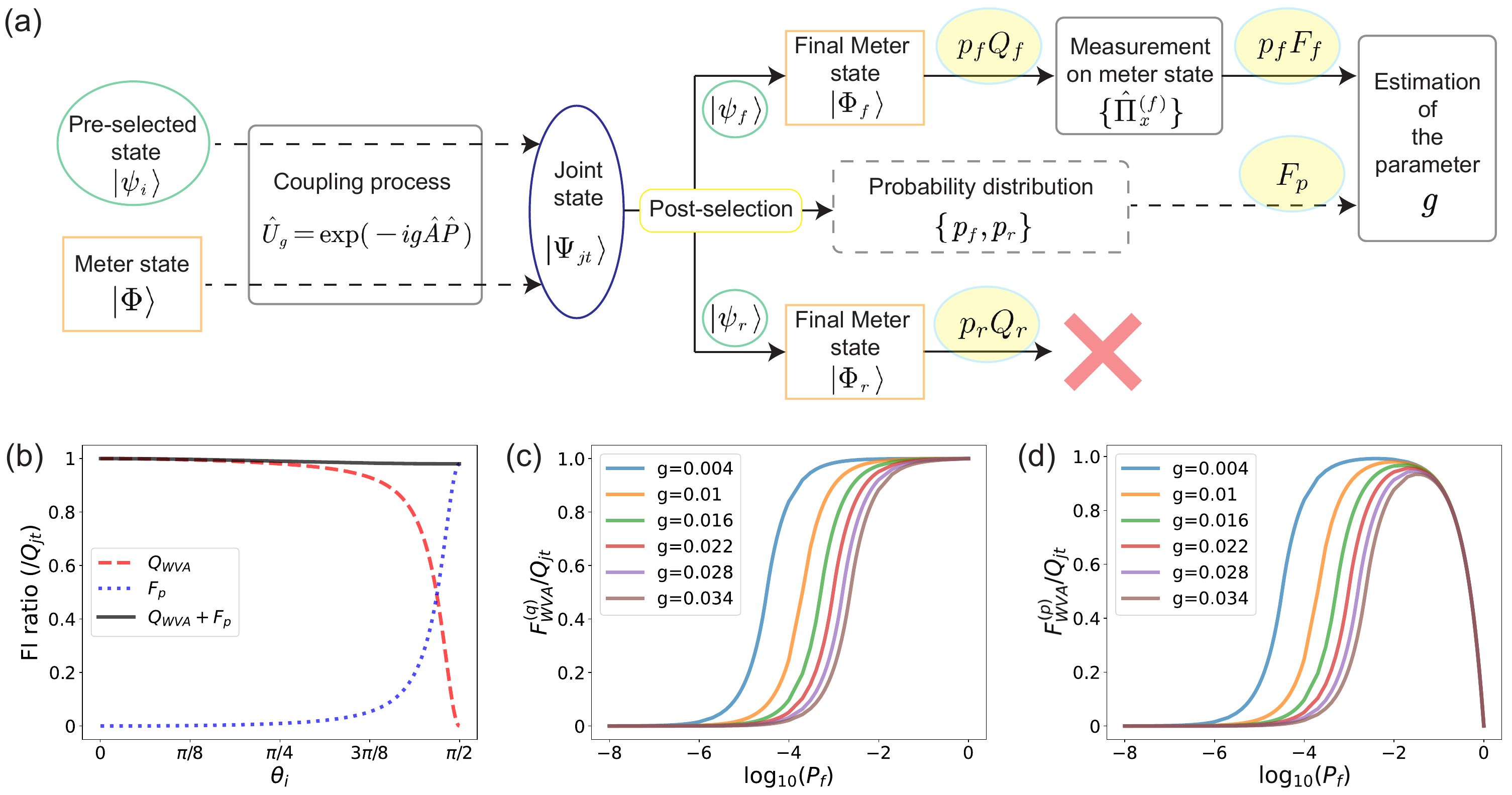}
	\caption{(a) The distribution of the FI during the post-selection process in WVA. (b) The ratio of the QFI of WVA $Q_{\text{WVA}}$ to $Q_{jt}$ and the ratio of the classical FI $F_p$ to $Q_{jt}$ are compared while varying the parameter $\theta_i$. The post-selected state keeps the optimal choice $|\psi_f^{\text{opt}}\rangle$ and $g/(2\sigma)=0.1$. (c) and (d) The maximum FI of the real and the imaginary WVA, $p_fF_f^{(q)}$ and $p_fF_f^{(p)}$, over each probability of post-selection are compared with different $g$ and $\sigma = 0.5$.}
	\label{fig:wva_metro}
\end{figure*}

The schematic diagram for analysing the ultimate precision of WVA is presented in Fig. \ref{fig:wva_metro} (a). The parameter $g$ is encoded into the joint state $|\Psi_{jt}\rangle$ through the interaction between the QS and the MS. By substituting the Eq. \eqref{eq:joint_state} ($|\Psi_{jt}\rangle$) into Eq. \eqref{eq:QFI_pure}, we obtain the QFI of the joint state 
\begin{eqnarray}
Q_{jt} & = & 4\Big[\langle \hat{A}^2\rangle_s \langle \hat{P}^2\rangle_m - \langle \hat{A}\rangle_s^2 \langle \hat{P}\rangle_m^2 \Big] {}\nonumber\\
& = & 4 \Big[ \langle \hat{A}^2\rangle_s \text{Var}_m (\hat{P}) + \text{Var}_s (\hat{A}) \langle \hat{P}\rangle_m^2 \Big] {}\nonumber\\
& = & 4\Big[ \text{Var}_s (\hat{A}) \langle \hat{P}^2\rangle_m +\langle \hat{A}\rangle_s^2 \text{Var}_m (\hat{P})\Big],
\label{eq:QFI_joint}
\end{eqnarray}
where $\text{Var}(\hat{O}) = \langle \hat{O}^2\rangle - \langle \hat{O}\rangle^2 $ is the variance of the operator $\hat{O}$ and $\langle \cdot \rangle$ denotes the average value in the initial state of the QS (subscript $s$) or the MS (subscript $m$). If the initial state of the QS (MS) is a balanced state, i.e., $\langle \hat{A}\rangle_s = 0$ ($\langle \hat{P}\rangle_m=0$), the QFI of the joint state can be simplified as $Q_{jt} = \text{Var}_s(\hat{A})\langle \hat{P}^2\rangle_m$ ($Q_{jt} = \langle \hat{A}^2\rangle_s\text{Var}_m (\hat{P})$). Additionally, both $|\psi_i\rangle$ and $|\Phi\rangle$ are balanced states, resulting in $Q_{jt} = 4\text{Var}_s(\hat{A})\text{Var}_m (\hat{P})$. The subsequent post-selection of the QS can be regarded as part of the measurements on the joint state, which cannot enhance QFI \cite{zhang2015precision}. Thus, the QFI of the joint state $Q_{jt}$ sets the upper bound for the measurement precision in the WVA scheme. Furthermore, the expression of $Q_{jt}$ also provides the possible approaches to improve the ultimate precision of WVA.

To account for the metrological contributions of all components during the post-selection process, we describe the post-selection by performing a POVM $\{|\psi_f\rangle\langle \psi_f|\otimes \hat{I}_m, |\psi_r\rangle\langle \psi_r|\otimes \hat{I}_m\}$ ($\langle \psi_r|\psi_f\rangle =0$) measurement on the joint state. This post-selection generates the success and failure modes of the final meter states $|\Phi_f\rangle$ and $|\Phi_r\rangle$, corresponding to the post-selection probabilities $p_f$ and $p_r = 1-p_f$, respectively. By substituting the expression of $|\Phi_f\rangle$ into Eq. \eqref{eq:QFI_pure}, we obtain the QFI of the post-selected MS $|\Phi_f\rangle$ with respect to $g$:
\begin{equation}
Q_f = 4\Big[ \langle \hat{J}^\dagger \hat{J}\rangle_m - \Big|\langle \hat{J}^\dagger \hat{K}\rangle_m  \Big|^2 \Big].
\label{eq:QFI_of_Qf}
\end{equation}
Here, we denote $\hat{J} =\langle \psi_f |\hat{A}\hat{P}\exp(-ig\hat{A}\hat{P})|\psi_i\rangle/\sqrt{p_f}$ and $\hat{K} = \langle \psi_f |\exp(-ig\hat{A}\hat{P})|\psi_i\rangle/\sqrt{p_f}$ \cite{alves2015weak}. The QFI $Q_r$ of $|\Phi_r\rangle$ can be derived similarly by replacing $|\Phi_f\rangle$ and $p_f$ with $|\Phi_r\rangle$ and $p_r$, respectively. Furthermore, the probability distribution $\{p_f,p_r\}$ in the post-selection process also contains information about $g$ quantified by the classical FI
\begin{equation}
F_p = \frac{1}{p_f(1-p_f)}\Big(\frac{dp_f}{dg} \Big)^2.
\end{equation}
All three contributions constitute the QFI of the joint state $Q_{jt} = p_fQ_f + p_rQ_r + F_p$ \cite{zhang2015precision}. Since the WVA scheme typically focuses on the amplified outcomes in the post-selected MS $|\Phi_f\rangle$, we regard $Q_{\text{WVA}} = p_fQ_f$ as the QFI of WVA. 

In order to realize the optimal precision of WVA, it is necessary to maximize the total QFI of the joint state $Q_{jt}$ by optimizing the pre-selected state. Subsequently, the post-selected state is optimized to make $Q_{\text{WVA}}$ attain the maximum $Q_{jt}$. As a demonstration, we employ the zero-mean Gaussian superposition state $|\Phi\rangle$ from Eq. \eqref{eq:initial_MS} as the MS and $|\psi_i\rangle$ ($|\psi_f\rangle$) serving as the pre- (post-) selected state in the QS. By considering the interaction Hamiltonian $\hat{H} = g\delta(t-t_0)\hat{A}\hat{P}$, we find that $Q_{jt} = \langle \psi_i|\hat{A}^2|\psi_i\rangle /\sigma^2$. When AAV approximate condition Eq. \eqref{eq:AAV_condition} is satisfied, the QFI of WVA is approximated as $Q_{\text{WVA}} \approx |\langle \psi_i|\hat{A}|\psi_f\rangle|^2/\sigma^2$. The maximal QFI of WVA $Q_{\text{WVA}}^{\text{max}}$ can achieve $Q_{jt}$ provided an optimal post-selected state $|\psi_f^{\text{opt}}\rangle = \hat{A}|\psi_i\rangle/\langle \hat{A}^2\rangle^{1/2}$. Consequently, the post-selected MS with a probability of $p_f = |\langle \psi_f^{\text{opt}}|\psi_i\rangle|$ can capture all the metrological information of the joint state. However, the probability of post-selection cannot be extremely small due to the AAV approximate condition. When $|\langle \psi_f^{\text{opt}}|\psi_i\rangle|$ becomes comparable to $g/\sigma$, $Q_{\text{WVA}}$ begins to decrease. For the case of $g|\langle \hat{A}\rangle_w|/\sigma \gg 1$, we find that $Q_{\text{WVA}}^{\text{opt}} \approx 0$ and $F_p^{\text{opt}} \approx Q_{jt}$. In this condition, all the metrological information about $g$ can be extracted from the statistics of the post-selection. The ratios of $Q_{\text{WVA}}/Q_{jt}$ and $F_p/Q_{jt}$ with respect to the angle $\theta_i$ of the pre-selected state $|\psi_i\rangle$ and the optimal post-selected state $|\psi_f^{\text{opt}}\rangle$ are depicted in Fig \ref{fig:wva_metro} (b).

In the preceding discussion, we focused on condensing all the QFI of the joint state into that of WVA, which we denote as $Q_{\text{WVA}} = Q_{jt}$. In the following, we investigate appropriate measurement strategies that extract the full QFI $Q_f$ about $g$ to the classical FI $F_f$ of the measured probability distributions. For a general weak value in standard WVA, we have derived the optimal observable $\hat{S}_\theta$ in section 2.1, which yields maximal shifts of the post-selected MS. It has been proven that this observable is also optimal for obtaining the maximal FI $Q_{f} = F_f^{(s_\theta)}$ \cite{knee2018seeing}. Here, we take two prevalent WVA schemes, namely, the real and imaginary WVA, for comparison. For the real (imaginary) WVA, the optimal parameters of the QS satisfying $\theta_i=-\theta_f$, $\phi = \phi_i-\phi_f=0$ ($\theta_i=-\theta_f=\pi/2$, $\phi\rightarrow 0$) yields $Q_{\text{WVA}}\approx Q_{jt}$. In the real (imaginary) WVA, we have the optimal observable $\hat{S}_\theta = \hat{Q}$ ($\hat{S}_\theta = \hat{P}$), from which we obtain the classical FI $p_fF_f^{(q)} = 1/\text{Var}(\hat{Q})$ ($p_fF_f^{(p)} = 4\text{Var}(\hat{P})$). Since the Gaussian MS has the minimum uncertainty, i.e., $\text{Var}_m (\hat{Q}) \text{Var}_m (\hat{P}) = \sigma^2 \cdot 1/(2\sigma)^2 = 1/4$, the maximal FI of both the real and the imaginary WVA are equal to the QFI of WVA $p_fF_f^{(q)} = p_fF_f^{(p)} = Q_{\text{WVA}}$. The difference lies in that the maximal FI of the real WVA can be attained with large range of weak values while the imaginary WVA achieves the maximal FI with only large modulus of weak values. In Fig. \ref{fig:wva_metro} (c) and (d), we compare the maximal FI of the real and imaginary WVA for each probability of post-selection $p_f$. The results indicate that the maximal FI of the real WVA $p_fF_f^{(q)}$ monotonically increases with a larger $p_f$, and $p_f=1$ corresponds to the FI of the CM $F_{\text{CM}}=Q_{\text{WVA}}$. In contrast, the maximal FI of the imaginary WVA $p_fF_f^{(p)}$ approaches $Q_{\text{WVA}}$ only with a small $p_f$. The adjustable success probability of post-selection without discarding the metrological information, provides the real WVA with the potential advantages in resisting noise and saturation of the detectors.

The real WVA has demonstrated advantages over the imaginary WVA in the adjustable range of $p_f$ for achieving the maximal FI. However, both the FI of the imaginary WVA $p_fF_f^{(p)}$ and $Q_{jt}$ are proportional to the variance of the observable $\hat{P}$, indicating the significant potential of the imaginary WVA. In certain scenarios, the uncertainty of the initial MS, denoted by the subscript $m^\prime$, is not at its minimum due to factors such as the generation of the MS (e.g., the squeezed state) or experimental noise, leading to $\text{Var}_{m^\prime}(\hat{Q})\text{Var}_{m^\prime}(\hat{P})>1/4$. The influence of noise on the ultimate precision of WVA will be discussed in sections 4 and 5. Here, we specifically consider the squeezed state with the observables $\hat{P}$ and $\hat{Q}$ referring to the quadratures in phase space. In cases where $\text{Var}_{m^\prime}(\hat{Q})<\text{Var}_{m}(\hat{Q})$ and $\text{Var}_{m^\prime}(\hat{P})>\text{Var}_{m}(\hat{P})$, both the maximal FI of the real WVA and the CM increase to $1/\text{Var}_{m^\prime}(\hat{Q})$. Notably, the imaginary WVA can leverage the variance of the observable $\hat{P}$ to achieve a larger FI of $4\text{Var}_{m^\prime}(\hat{P})$. Considering that $Q_{jt}\ge 4\Delta^2_{m^\prime}\hat{P}$, the imaginary WVA offers an optimal measurement strategy for extracting the full QFI of the joint state. In contrast, the FI of the real WVA is smaller than that of the imaginary WVA due to suboptimal pre- and post-selection as well as the measurement on the final MS. This property of the imaginary WVA presents an avenue for enhancing the precision of WVA by introducing squeezing, mixture and even noise into the MS, which will be further discussed in the subsequent two sections.

\section{4. Weak-value amplification with Heisenberg-limited precision}

\begin{figure*}[t]
	\centering
	\includegraphics[width=0.95\textwidth]{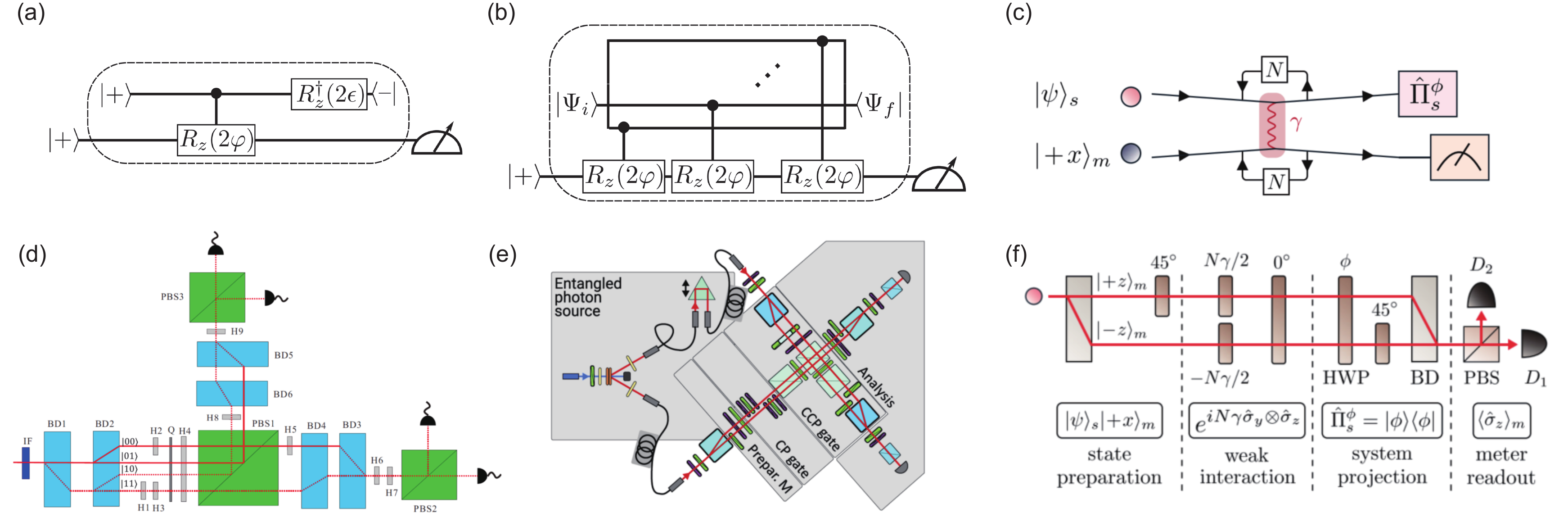}
	\caption{(a) The quantum circuit for implementing the standard WVA. (b) The quantum circuit for realizing the quantum-enhanced WVA \cite{pang2014entanglement}. (c) WVA with iterative interactions \cite{kim2022heisenberg}. (d) Realization of entanglement-assisted WVA with single-photon simulation \cite{chen2019realization}. (e) Two-photon entangled state demonstration for realizing the entanglement-assisted WVA \cite{starek2020experimental}. (f) Experimental setup for iterative-interaction WVA \cite{kim2022heisenberg}.}
	\label{fig:entangleWVA}
\end{figure*}

Considerable research has been conducted regarding the ultimate precision of standard WVA, which has prompted researchers to investigate its potential to surpass the SQL , and achieve Heisenberg-limited (HL) precision. In recent years, various enhanced WVA schemes capable of achieving Heisenberg-scaling precision have been reported. These advancements are realized through two approaches: the first involves exploiting entangled states or iterative interactions, drawing inspiration from the phase estimation scheme in standard quantum metrology. The second approach relies on the phase-space interaction scenario, which enables the attainment of Heisenberg-scaling precision using solely classical resources, in a way similar to nonlinear quantum metrology \cite{boixo2007generalized, boixo2008quantum}.

\subsection{4.1 Heisenberg-limited weak-value amplification in linear optical system}

We first introduce a phase-estimation WVA scheme with its corresponding schematic diagram shown in Fig. \ref{fig:entangleWVA} (a). The pre-selected QS and the initial MS are both qubit states $|+\rangle_s  = |+\rangle_m = (|0\rangle+|1\rangle)/\sqrt{2}$, with the control gate $R_z(2\varphi) = \exp(-i\varphi \hat{\sigma}_z^{(s)}\otimes \hat{\sigma}_z^{(m)}/2)$ coupling them, resulting in a joint state of $|\Psi_{b}\rangle = R_z(2\varphi) |+\rangle_s|+\rangle_m$. Subsequent post-selection on the QS by $R_z(2\epsilon)|-\rangle = (e^{-i\epsilon}|0\rangle - e^{i\epsilon}|1\rangle)/\sqrt{2}$ amplifies the parameter $\varphi$ by the imaginary weak value $\langle \hat{A}\rangle_w = i/\epsilon$ ($\epsilon \ll 1$), leading to a shift of the observable $\hat{\sigma}_z$ on the final MS, given by $\langle\hat{\sigma}_z\rangle_f = \varphi \text{Im}\langle \hat{A}\rangle_w$. Evaluating the ultimate precision of $\varphi$ using the FI metric, we find that the QFI about $\varphi$ of the joint state is $Q_{b} = 4\text{Var}_+(\hat{A}) \text{Var}_+(\hat{\sigma}_z) = 4$. Considering the probability of post-selection $p_f \approx \epsilon^2$, the classical FI of measuring the observable $\hat{\sigma}_z$ on $|+^\prime\rangle$ is $F_{\text{WVA}} = p_fF_f = 4$. This result shows that both the post-selected state and the measurement observable on the MS are optimal for concentrating the full QFI into a fraction of post-selected MS. For $N$ independent trials of WVA, the average number of successful post-selection is $n = 1-(1-p_f)^N\approx N\epsilon^2$, which scales linearly with $N$. Therefore, the total FI of all $N$ independent measurements is given by $F_{\text{WVA}}^{(N)} =nF_f= 4N$, showing a precision in SQL.

In 2014, Pang et. al. proposed an entanglement-assisted WVA scheme that improves the measurement precision of $\varphi$ to the HL \cite{pang2014entanglement}. The quantum circuit for this scheme is depicted in Fig. \ref{fig:entangleWVA} (b), where $N$ quantum-correlated particles form the initial state of the QS $|\Psi_i\rangle$. The interaction between the QS and the MS is described by the unitary transformation $\hat{U} = \exp(-i\varphi \hat{A}^{(N)}\hat{\sigma}_z^{(m)}/2)$, in which the observable of the QS becomes $\hat{A}^{(N)} = \sum_k \hat{A}_k$ with $\hat{A}_k = \hat{I}\otimes \cdots \hat{\sigma}_z^{(s)} \otimes \cdots \otimes \hat{I}$ referring to the observable $\hat{\sigma}^{(s)}_z$ acting on $k$th state. The interaction results in the joint state $|\Psi_{jt}^{(N)}\rangle$. Assuming that $|\Psi_i\rangle$ is a balanced state ($\langle \Psi_i| \hat{A}|\Psi_i\rangle_s = 0$), the QFI of $|\Psi_{jt}^{(N)}\rangle$ about the parameter $\varphi$ is given by $Q_{jt}^{(N)} = 4\text{Var}_{|\Psi_i\rangle}[\hat{A}^{(N)}] \text{Var}_+(\hat{\sigma}^{(m)}_z) = 4 \text{Var}_{|\Psi_i\rangle}[\hat{A}^{(N)}]$. Thus, the maximal QFI $[Q_{jt}^{(N)}]_{\text{max}} = 4\{\text{Var}_{|\Psi_i\rangle}[\hat{A}^{(n)}]\}_{\text{max}}=4N^2$ can be achieved by preparing the total pre-selected state $|\Psi_i\rangle$ as the maximally entangled form 
\begin{equation}
|\Psi_i^{(e)}\rangle = \frac{1}{\sqrt{2}}(|0\rangle^{\otimes N} + e^{i\theta}|1\rangle^{\otimes N})
\end{equation}
with $\theta$ an arbitrary relative phase. In this scheme, the QFI in the entanglement-assisted WVA $Q_{jt}^{(N)}$ indicates the Heisenberg-scaling precision and shows an improvement over that of the standard WVA by a factor of $N$.

In order to fully extract the QFI of the joint state, Pang et al. proposed two optimizations for the post-selected state: (i) maximizing the probability of post-selection $p_f^{(N)}$,while keeping the weak value $\langle \hat{A}^{(N)}\rangle_w$ fixed; (ii) maximizing the modulus of the weak value $|\langle \hat{A}^{(N)}\rangle_w|$ while fixing $p_f^{(N)}$. In the former case, the weak value is set as $\langle \hat{A^{(N)}}\rangle_w = i/\epsilon$. By optimizing the post-selected state, the probability of post-selection can be maximized to $p_f^{(N)} = N^2\epsilon^2$, which is achieved by the following expression:
\begin{equation}
|\Psi_f^{(N)}\rangle_p \propto (N+\langle \hat{A}^{(N)}\rangle_w^*)|0\rangle^{\otimes N} + (N-\langle \hat{A}^{(N)}\rangle_w^*)|1\rangle^{\otimes N} \propto e^{-in\epsilon}|0\rangle^{\otimes N} - e^{in\epsilon}|1\rangle^{\otimes N}.
\end{equation}
In the latter case, the probability of post-selection is set as $p_f^{(N)} = N\epsilon^2$, which is equivalent to $Np_f$ in the standard WVA. The post-selected state can be decomposed as follows: 
\begin{equation}
|\Psi_f^{(N)}\rangle = \sqrt{p_f^{(N)}}|\Psi_i\rangle +\sqrt{1-p_f^{(N)}}|\Psi_i^{\perp}\rangle,
\end{equation}
where $|\Psi_i^{\perp}\rangle$ is orthogonal to $|\Psi_i\rangle$. The maximal modulus of the imaginary weak value can be approximated as:
\begin{equation}
\text{max} |\langle \hat{A}^{(N)}\rangle_w| \approx \sqrt{\text{Var}_{|\Psi_i^{(e)}\rangle}[\hat{A}^{(N)}]/p_f^{(N)}} = \sqrt{N}/\epsilon.
\end{equation}
This maximum value is achieved when $|\Psi_i^{\perp}\rangle$ is parallel to the component of $\hat{A}^{(N)}|\Psi_i\rangle$ in the complementary subspace orthogonal to $|\Psi_i\rangle$. Therefore, the optimal post-selected state for maximizing the weak value is given by 
\begin{equation}
|\Psi_f\rangle_w \propto e^{-i\sqrt{N}\epsilon}|0\rangle^{\otimes N} - e^{i\sqrt{N}\epsilon}|1\rangle^{\otimes N}.
\end{equation}
Since the QFI of WVA $Q_{\text{WVA}}^{(N)}$ is equal to $Q_{jt}^{(N)}$ in both cases, both the post-selected states $|\Psi_f^{(N)}\rangle_p$ and $|\Psi_f^{(N)}\rangle_w$ are optimal for concentrating the full QFI $Q_{jt}^{(N)}$ into the post-selected MS. Furthermore, the QFI of entanglement-assisted WVA can be extracted by measuring the optimal observable $\hat{\sigma}_z^{(m)}$ on the final MS.

Theoretical analysis has shown that the entanglement-assisted WVA scheme is advantageous in addressing the inaccuracies in measurement results and the loss of FI caused by readout errors \cite{pang2015suppressing}. In 2019, Chen et al. have conducted an experimental demonstration of the enhanced weak values or the post-selection probabilities obtained through entanglement-assisted WVA using the high-dimensional DOF of single photons shown in Fig. \ref{fig:entangleWVA} (d)\cite{chen2019realization}. Furthermore, in 2020, a two-photon entanglement-assisted WVA was realized with the experimental setup illustrated in Fig. \ref{fig:entangleWVA} (e) \cite{starek2020experimental}. The experimental findings revealed that the precision improvement resulting from entanglement is highly sensitive to the purity of the QS state. Due to limitations in current photonic technology, achieving the theoretical Heisenberg scaling of the entanglement-assisted WVA remains a challenge. 

To circumvent this technical obstacle, Kim et al. proposed a WVA scheme based on iterative interactions, which achieves the HL without the use of entangled resources \cite{kim2022heisenberg}. The schematic diagram of the iterative WVA and its experimental implementation are illustrated in Fig. \ref{fig:entangleWVA} (c) and (f), respectively. In this scheme, the QS is represented by the spin of photons with a pre-selected state $|\psi\rangle_s = (|+y\rangle_s +|-y\rangle_s)/\sqrt{2}$, where $|+y\rangle_s$ ($|+y\rangle_s$) refers to the right (left) circular polarization state. The optical path state serves as the MS and is initialized as the superposition state $|+x\rangle_m = (|+z\rangle_m + |-z\rangle_m)/\sqrt{2}$, where $|+z\rangle_m$ and $|-z\rangle_m$ represent the upper and lower path states, respectively. The unitary transformation for $N$ iterative interactions is described by $\hat{U} = e^{iN\gamma \hat{A}\otimes \hat{\sigma}_z}$, where $\gamma$ is the coupling strength and $\hat{A}$ ($\hat{\sigma}_z$) represents the observable of the QS (MS). The post-selected states have the same probability amplitudes as those in $|\Psi_f^{(N)}\rangle_p$ and $|\Psi_f^{(N)}\rangle_p$, but the base states $|0\rangle^{\otimes N}$ and $|1\rangle^{\otimes N}$ are replaced by $|+y\rangle_s$ and $|-y\rangle_s$, respectively. Correspondingly, we denote the post-selected states as
$|\psi_f^{(iter)}\rangle_p \propto  e^{-iN\epsilon}|+y\rangle_s + e^{iN\epsilon}|-y\rangle$ and $|\psi_f^{(iter)}\rangle_w \propto   e^{-i\sqrt{N}\epsilon}|+y\rangle_s + e^{i\sqrt{N}\epsilon}|-y\rangle$. Similarly to entanglement-assisted WVA, the post-selected state $|\psi_f^{(iter)}\rangle_p$ ($|\psi_f^{(iter)}\rangle_w$) improves the probabilities of post-selection $p_f^{(iter)}$ (weak value $\langle N\hat{\sigma}_y\rangle_w$) by a factor of $N$ ($N$) compared to standard WVA. In both cases, the QFI of the iterative WVA $Q_{\text{WVA}}^{(iter)} = N^2$, achieves Heisenberg-scaling precision. Compared to entanglement-assisted WVA, the iterative scheme does not suffer from the fragility and the preparation of entangled resources, thus providing a practical pathway for Heisenberg-scaling WVA. However, the interaction time of the iterative WVA is $N$ times larger than that of the entanglement-assisted WVA for achieving the same level of precision.

\subsection{4.2 weak-value amplification with phase-space interaction}

\begin{figure*}[t]
	\centering
	\includegraphics[width=0.9\textwidth]{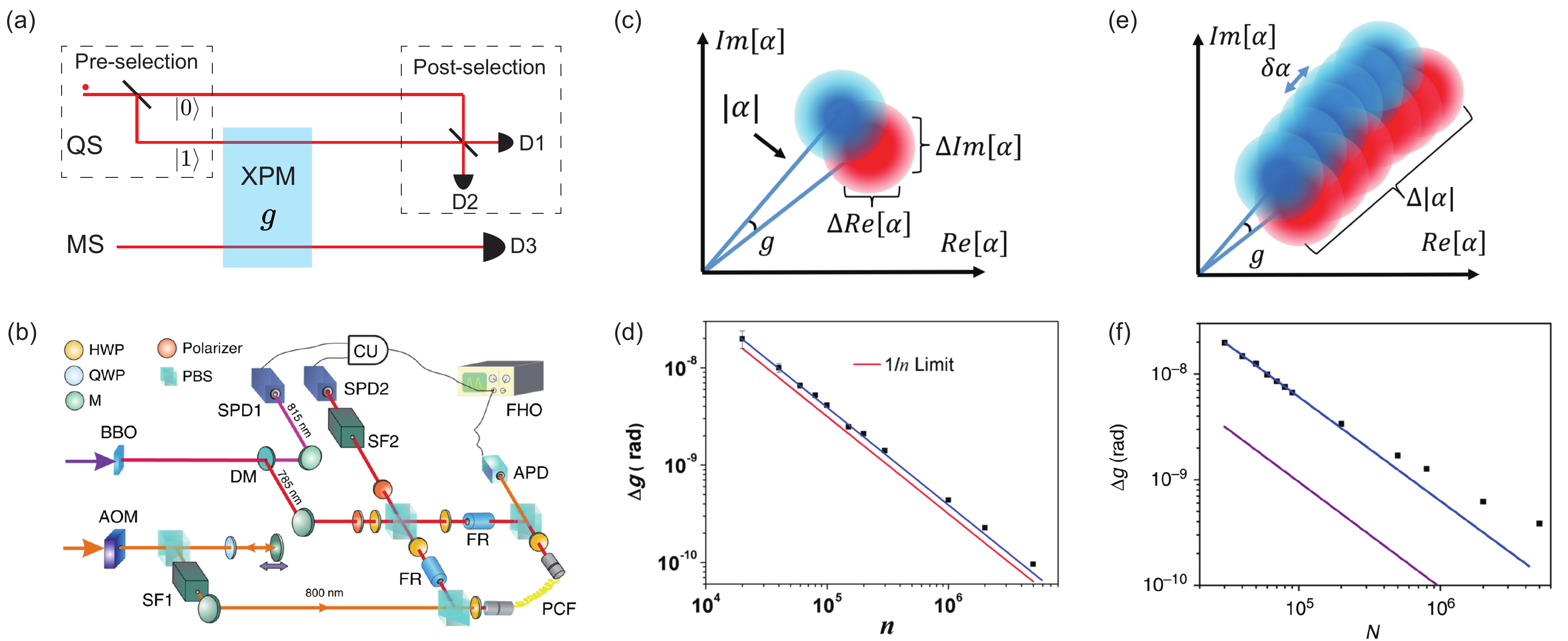}
	\caption{(a) The schematic diagram for the WVA of phase-space interaction. (b) The experimental setup to implement the phase-space WVA through the Kerr effect \cite{chen2018heisenberg}. (c) Uncertainty of the coherent state $|\alpha\rangle$ to estimate the coupling strength $g$ \cite{chen2021beating}. (d) The precision of $g$ with the average photon number of coherent state $N$ \cite{chen2018achieving}. (e) Uncertainty of the mixed coherent state to estimate the coupling strength $g$ \cite{chen2021beating}. (f) The precision of $g$ with the average photon number of mixed coherent state $n$ \cite{chen2018heisenberg}.} 
	\label{fig:phasewva}
\end{figure*}

The phase-space distribution provides a comprehensive representation of the quantum statistical characteristics of photonic states. The interaction in phase space can alter the photon-number distribution. In 2015, Zhang et al. considered the phase-space WVA scheme with the schematic diagram depicted in Fig. \ref{fig:phasewva} (a) \cite{zhang2015precision}. The which-path information (up and down) of single photons in the Mach-Zehnder interferometer is referred to as the QS. The pre-selected state of the QS is parametrized as $|\psi_i\rangle = \cos(\theta_i/2)|0\rangle + \sin(\theta_i/2)|1\rangle$, where $|0\rangle$ and $|1\rangle$ denote the number of photons at the downward arm participating in the interaction. For simplicity, we consider a photonic state $|\Phi_p\rangle$ as the MS. The cross-phase modulation (XPM) between the QS and the MS is described by the Hamiltonian $\hat{H}_p = g\delta(t-t_0)\hat{C}\hat{n}_p$, where $\hat{C} = |1\rangle\langle 1|$ is the observable of the QS and $\hat{n}_p$ is the photon-number operator of the MS. The small interaction strength $g$ specifies the third-order nonlinear coefficient, which is the unknown parameter to be measured. The interaction produces a weakly entangled joint state, given by 
\begin{equation}
|\Psi_{jt}^{(p)}\rangle = \cos(\theta_i/2) |0\rangle |\Phi_p\rangle + \sin(\theta_i/2) |1\rangle e^{ig\hat{n}_p}|\Phi_p\rangle.
\end{equation}
According to Eq. \eqref{eq:QFI_joint}, the QFI of the joint state about $g$ is given by $Q_{jt}(|\Phi_p\rangle) = 4[\langle \hat{C}^2\rangle_s\text{Var}_p(\hat{n}) + \text{Var}_s(\hat{C})\langle \hat{n}\rangle_p^2]$. By setting $\theta_i = \pi/2$, the QFI can be simplified as $Q_{jt}(|\Phi_p\rangle) = 2\text{Var}_m(\hat{n})+\langle \hat{n}\rangle_m^2$. Therefore, the photon-number statistics of the MS affects how the QFI scales with the mean photon numbers. This feature offers possible approaches to surpass the SQL or achieve the HL with classical photonic states in phase-space WVA. 

In 2015, Zhang et al. proposed a scheme to use the coherent state $|\alpha\rangle$ as the meter state \cite{zhang2015precision}. When $|\alpha| \gg 1$, the term associated with the mean photon number $\langle \hat{n}\rangle_m^2$ dominates $Q_{jt}$ and exhibits the Heisenberg-scaling of the ultimate precision. In the work, a comprehensive analysis of the FI distribution during the post-selection process is established. As shown in Fig. \ref{fig:wva_metro}, the post-selection divides the $Q_{jt}$ into three components $Q_{jt} = p_fQ_f + p_rQ_r + F_p$. In weak-measurement regime ($g\ll 1$), we can approximate the results as follows: $p_fQ_f \approx  (1-\epsilon^2/4)N$, $p_rQ_r \approx \epsilon^2 N/4$, $F_p \approx N^2$, and the success probability $p_f = [1-\cos(2gN+\epsilon)]/2$. These results suggest that the Heisenberg-scaling precision lying in $F_p$ can be achieved by observing the probability distribution $\{p_f,p_r\}$. Jordan et. al. also provided a simple interpretation of the Heisenberg-scaling precision and an approach to extracting the information from the perspective of state discrimination \cite{jordan2015heisenberg}. As Fig. \ref{fig:phasewva} (f) shows, Chen et. al. have achieved practical Heisenberg-scaling precision, with $\Delta g \approx 1.2/n$ up to $10^{-10}$ rad, with the mean photon number over $10^6$.

Next, we consider the metrological contribution of the variance term $\text{Var}_p(\hat{n})$ to the total QFI $Q_{jt}$. With a coherent MS $|\alpha\rangle$, the measurement of the post-selected MS obtains the FI $p_fQ_f \approx  (1-\epsilon^2/4)N$, only achieving the ultimate precision of $g$ in SQL. However, the variance $\text{Var}_p(\hat{n})$ can be easily increased through classical operations. Moreover, both the phase-estimation experiment using white light and the theoretical derivation by Kedem have indicated that the variance of incoherent MS or due to classical fluctuations equivalently contributes to the amplification factors multiplied by the imaginary weak values \cite{xu2013phase, kedem2012using}. Motivated by this observation, Chen et al. utilized a mixture of coherent states as the initial MS: $\rho_p = \sum_j p_j |\alpha_j\rangle\langle \alpha_j|$, with the phase-space distribution shown in Fig. \ref{fig:phasewva} (e) \cite{chen2018achieving}. Given the pre- and post-selected states $|\psi_i\rangle$ and $|\psi_f\rangle$, along with the interaction Hamiltonian $\hat{H}_p$, the final MS evolves according to
\begin{equation}
\rho_p^\prime = \frac{1}{\sqrt{\mathcal{N}_p}}\langle \psi_f|\exp(-ig\hat{C}\hat{n}_p)|\psi_i\rangle\langle \psi_i|\otimes \rho_p \exp(ig\hat{C}\hat{n}_p)|\psi_f\rangle,
\end{equation}
where the normalization coefficient $\mathcal{N}_p$ represents the success probability of post-selection. Thus, the change in the average photon number of the final MS can be approximated as:
\begin{equation}
\text{Tr}(\rho_p^\prime \hat{n}_p) - \text{Tr}(\rho_p \hat{n}_p) \approx 2g \text{Im}\langle \hat{C}\rangle_w \text{Var}_p(\hat{n}_p),
\end{equation}
where $\langle \hat{C}\rangle_w$ denotes the weak value and $\text{Var}_p(\hat{n}_p) = \text{Tr}(\rho_p \hat{n}_p^2) - (\text{Tr}(\rho_p\hat{n}_p))^2$. The overall classical FI about $g$ in phase-space WVA can be obtained by substituting the photon-number statistics $f(n,g) = \langle n|\rho_p^\prime|n\rangle$ into Eq. \eqref{eq:FI_def}:
\begin{equation}
F_{\text{WVA}}^{(ps)} = 4 \mathcal{N}_p \text{Im}\langle \hat{C}\rangle_w^2 \text{Var}_p(\hat{n}_p).
\end{equation}
In their experiment, the pre- and post-selected states were chosen as $|\psi_i\rangle = (|0\rangle + |1\rangle)/\sqrt{2}$ and $|\psi_i\rangle = (|1\rangle - e^{-i\epsilon}|0\rangle)/\sqrt{2}$ with $\epsilon=0.1$. The mixture of coherent states maintains $\text{Var}_p(\hat{n}_p) = N^2/4$ to achieve Heisenberg-scaling precision for values of $N$ ranging from $3\times 10^4$ to $5\times 10^6$. According to the CRB, the ultimate precision of $g$ for $\nu = 2.2\times 10^5$ measurements is given by $\text{Var}(g) = 1/(\nu F_{\text{WVA}}^{(ps)}) \approx 4 \times 10^{-7}/N^2$. As depicted in Fig. \ref{fig:phasewva} (d), the experimental precision of $g$ maintains Heisenberg-scaling precision up to $N=10^5$ photons and achieves an unprecedented precision $\Delta g = \sqrt{\text{Var}(g)} = 3.6\times 10^{-10}\text{rad}$.

\section{5. Advantages of weak-value amplification in the presence of practical imperfections}

In section 3 and 4, we have reviewed how quantum noise determines the ultimate precision of WVA and introduced approaches to surpass the SQL. However, experimental imperfections such as technical noise and nonideal detectors are inevitable factors that prevent measurement protocols from achieving the quantum-limited precision in practice. The amplification effect resulting from post-selection in WVA has been shown to be beneficial in resisting various types of technical noise. This noise tolerance has contributed to unprecedented measurement sensitivity in many experiments without requiring special noise-isolation operations. In this section, we review recent discussions about the metrological advantages of WVA over CM with the ubiquitous noise in optical systems. We try to construct a unified noise model for each case and analyze the metrics of SNR and FI. Our aim is to elucidate the essential mechanism that offers the WVA potential advantages under certain conditions.

\subsection{5.1 Weak-value amplification against time-correlated noise}   

\begin{figure*}[t]
	\centering
	\includegraphics[width=0.95\textwidth]{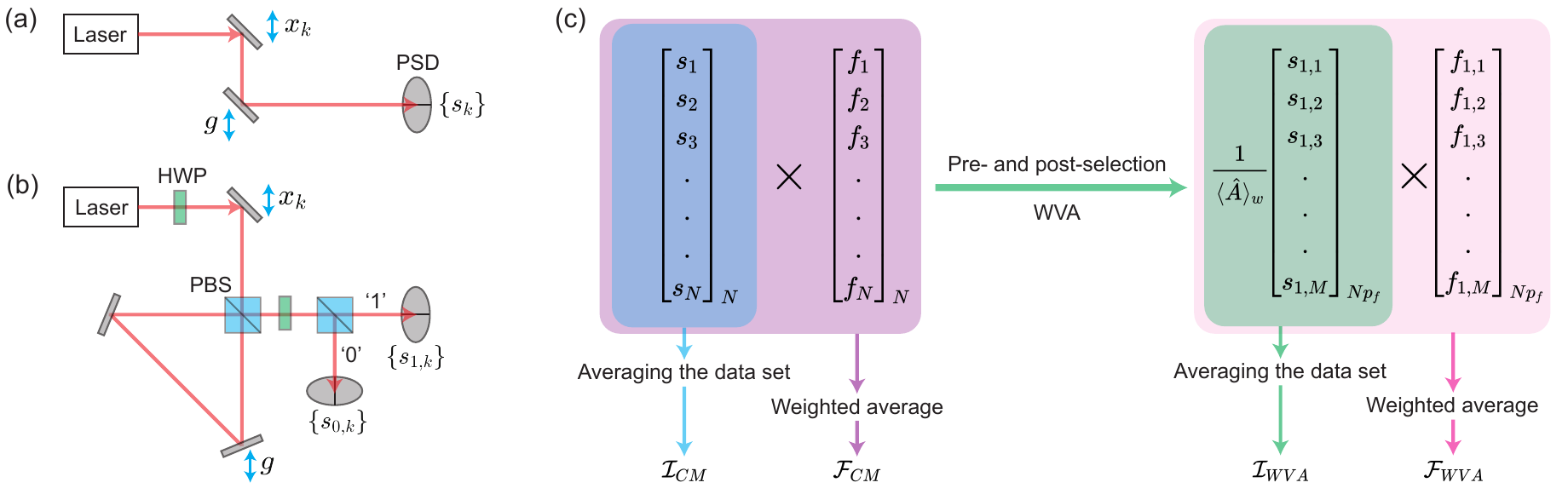}
	\caption{Comparison between WVA and CM in the presence of additive noise. (a) Measuring the displacement of a mirror $g$ with a random displacement $x_k$ in CM. (b)Measuring the displacement of a mirror $g$ inside the Sagnac interferometer with a random displacement $x_k$ in real WVA. (c) The data set used for estimating the parameter in CM and WVA is schematically illustrated.}
	\label{fig:corre_noise}
\end{figure*}

In 2011, Feizpour, Xing and Steinberg theoretically demonstrated the advantages of WVA over the CM in the presence of additive, time-correlated, Gaussian random noise with the SNR metric \cite{feizpour2011amplifying}. In 2014, Ferrie and Combes employed the FI metric to compare the precision of WVA and CM with the time-correlated noise. They found that the post-selection in WVA discarded the metrological information. Thus, WVA is suboptimal in precision compared to the CM with the optimal estimator \cite{ferrie2014weak}. Subsequently, Jordan et al. compared the FI metric with the SNR metric from a practical standpoint \cite{jordan2014technical}. In 2017, Sinclair et al. conducted comprehensive analysis on realistic models of time-correlated noise to clarify the conflicting claims about the optimal estimator and the usefulness of WVA \cite{sinclair2017weak}. We adopt a commonly used scenario for time-correlated noise to summarize the previous findings and highlight the metrological advantages of WVA.

Fig. \ref{fig:corre_noise} (a) and (b) illustrate practical setups to measure the displacement of a mirror using CM and WVA, respectively. The polarization and transverse spatial modes of photons are referred to as the QS and the MS, respectively. The additive time-correlated technical noise is modelled by the random displacement of the first mirror. In CM, measurements are typically repeated sequentially under the same circumstances to obtain a data set $\{s_k\}$, where $s_k = g + x_k$ refers to the $k$th measurement result and $x_k$ represents the zero-mean Gaussian noise term described by a covariance matrix $C_{k,l} = \langle x_k x_l\rangle$. As an experimentally relevant example, the covariance matrix can be modelled by
\begin{equation}
C_{k,l} = a \delta_{k,l} + c\exp(-|k-l|\delta t /\tau),
\end{equation}
where $a \delta_{k,l}$ represents a white-noise floor. In the time-correlated term $c\exp(-|k-l|\delta t /\tau)$, $\delta t$ and $\tau$ denote the time interval between two measurements and the correlation time of noise, respectively. We consider two limiting cases: when $\tau/\delta t \ll 1$, $C_{k,l} = (a+c)\delta_{k,l}$ represents the white-noise limit; in the opposite case, as $\tau/\delta t \gg 1$, $C_{k,l} = a\delta_{k,l} + c$ indicates the slow-noise limit. 

\begin{table}[]
\caption{The metrological information of WVA and CM with the SNR and FI metrics in different noise conditions}
\label{Table:correlated_noise}
\begin{tabular}{c|c|c|cc}
\hline
\makecell{Metrological \\ Information} & General & \makecell{White-noise limit \\($\tau \ll \delta t$)} & \multicolumn{1}{c|}{\makecell{Slow-noise limit 1 \\ ($\delta t \ll \tau$, $p_f \ll \delta t/\tau$)}} & \makecell{Slow-noise limit 2\\ ($\delta t \ll \tau$, $p_f \gg \delta t/\tau$) } \\ \hline
                       $\mathcal{I}_{\text{CM}}$   &  $\frac{1}{V}$  &         $\frac{N}{a+c}$          &   \multicolumn{2}{c}{ $\frac{N}{a+Nc}$} \\ \hline
         $\mathcal{F}_{\text{CM}}$  &    $\sum_{i,j}^N [C^{-1}]_{i,j}$     &                  $\frac{N}{a+c}$ &                      \multicolumn{2}{c}{$\frac{N}{a}$}     \\ \hline
         $\mathcal{I}_{\text{WVA}}$  &   $\frac{1}{V^\prime}$   &    $\frac{N}{a+c}$               & \multicolumn{1}{c|}{$\frac{N}{a+c}$}                &    $\langle \hat{A}\rangle_w^2 \frac{p_f N}{a + p_f Nc}$                 \\ \hline
       $\mathcal{F}_{\text{WVA}}$   &   $\sum_{i,j}^{Np_f} [C^{\prime -1}]_{i,j}$      &      $\frac{N}{a+c}$             & \multicolumn{1}{c|}{$\frac{N}{a+c}$}                   &        $\frac{N}{a}$            \\ \hline
\end{tabular}
\end{table}

In the following, we first examine the metrological performance of CM using both SNR and FI metrics. As discussed in section 2.1, the SNR metric corresponds to the estimator of AMR, denoted as $\hat{g}_{\text{AMR}}^{(\text{CM})} = \sum_{k=1}^N s_k/N$. The corresponding variance of AMR estimator is given by $V = \sum_{k,l}^N\langle s_k s_l\rangle/N^2$. Analogous to the FI, we define the metrological information of SNR as the inverse of the variance of AMR estimator, i.e., $\mathcal{I}_{\text{CM}} = 1/V$. Specifically, in the white-noise limit ($\delta t \gg \tau$), $\mathcal{I}_{\text{CM}} = N/(a+c)$. In contrast, in the slow-noise limit ($\delta t \ll \tau$), $\mathcal{I}_{\text{CM}} = N/(a+Nc)$. As $N$ tends to infinity, the information $\mathcal{I}_{\text{CM}}$ saturates at $1/c$ for typical positive correlations ($c>0$). This result is equivalent to an unknown offset ($c$) which cannot be reduced by large number of measurements.

To facilitate the derivation of the FI of $g$ in CM, we transform the measurement results $\{s_k\}$ into the vector form, denoted as $\vec{s}$. Consequently, we obtain the joint probability distribution of the measurement results $\vec{s}$ as a high-dimensional Gaussian distribution:
\begin{equation}
P_s(\vec{s}) = \frac{1}{\sqrt{(2\pi)^N\text{det} C}}\exp\Big[-\frac{(\vec{s}-\mu)C^{-1}(\vec{s}-\mu)}{2} \Big],
\end{equation}
where $C^{-1}$ denotes the inverse of the covariance matrix $C$. According to Eq. \eqref{eq:FI_def}, the FI regarding the parameter $g$ is given by
\begin{equation}
\mathcal{F}_{\text{CM}} = \sum_{k,l=1}^N [C^{-1}]_{k,l}.
\end{equation}
Consequently, the ultimate precision determined by the CRB can be asymptotically achieved with the maximum likelihood estimator
\begin{equation}
\hat{g}_{\text{MLE}}^{(\text{CM})} = \frac{\sum_{k,l=1}^N [C^{-1}]_{k,l}s_k}{\sum_{k^\prime,l^\prime=1}^N [C^{-1}]_{k^\prime l^\prime}} = \sum_k f_k s_k.
\label{eq:I_CM_correnoise}
\end{equation}
Here, this estimator can be interpreted as the weighted average of each measurement result $s_j$ with the factor $f_k = \sum_{l=1}^N [C^{-1}]_{k,l}/\sum_{k^\prime, l^\prime =1}^N [C^{-1}]_{k^\prime l^\prime}$. In comparison, the SNR method treats every data point equally with a weighting factor $f_k=1/N$. In the white-noise limit ($\delta t \gg \tau$), the maximum likelihood estimator $\hat{g}_{\text{MLE}}^{(\text{CM})}$ is equivalent to $\hat{g}_{\text{AMR}}^{(\text{CM})}$, resulting in $F_{\text{CM}} = \mathcal{I}_{\text{CM}} = N/(a+c)$. In the slow-noise limit ($\delta t \ll \tau$), the FI is approximated as $F_{\text{CM}} = \sum_{k,l} C^{-1}_{k,l} \approx  \sum_{k,l} [(a+cN)\delta_{kl}-c]/(a^2+Nac) \approx N/a$ for large $N$.

We denote the measurement results in WVA as $s_k^\prime = g\langle \hat{A}\rangle_w + x_k^\prime $, where the parameter $g$ is amplified by the weak value $\langle \hat{A}\rangle_w$ and $x_k^\prime$ represents the post-selected noise. Given the success probability of post-selection in WVA as $p_f$, the size of the post-selected signals is reduced to $N^\prime = Np_f$. Assuming a real weak value $\langle \hat{A}\rangle_w$, the optimal precision of real WVA satisfies the trade-off relation $p_f = \langle \hat{A}\rangle_w^2$. We denote the variance of the post-selected noise as $V^\prime = (1/{N^\prime}^2)\sum_{k,l}^{N^\prime} \langle x_k^\prime x_l^\prime\rangle $. Therefore, the metrological information of the parameter $g$ in WVA can be generally expressed by $\mathcal{I}_{\text{WVA}} = \langle \hat{A}\rangle_w^2 /V^\prime$. In the white-noise limit ($\delta t \gg \tau$), the information simplifies to $\mathcal{I}_{\text{WVA}} = N/(a+c)$, which is equal to $\mathcal{I}_{\text{CM}}$. Hence, WVA does not exhibit advantages over CM in the presence of additional white noise. As post-selection significantly alters the time correlation of the noise, we further divide the slow-noise limit into two cases based on the relationship between the post-selection probability $p_f$ and the ratio $\delta t/\tau$. When $p_f \ll \delta t/\tau$, the time correlation of post-selected measurements is close to white noise. In this situation, the metrological information is approximated as $\mathcal{I}_{\text{WVA}} = N/(a+c)$. In contrast, when $p_f \gg \delta t/\tau$, the noise in the post-selected results is still fully time-correlated, leading to the metrological information $\mathcal{I}_{\text{WVA}} = \langle \hat{A}\rangle_w^2 p_f N/(a + p_f Nc) \approx 1/(p_f c)$ for a large $N$. Compared to the metrological information of CM, the amplification effect in WVA improves the information from $\mathcal{I}_{\text{CM}} = 1/c$ to $\mathcal{I}_{\text{WVA}} = 1/(p_fc)$, which aligns with the findings in Ref. \cite{feizpour2011amplifying}.

In the table \ref{Table:correlated_noise}, we have summarized the metrological information of WVA and CM with both the SNR and FI metrics in various noise situations. In white-noise (uncorrelated) limit, the metrological information of AMR estimator is identical to the FI. The CM and WVA have identical FI, demonstrating an equivalent performance. However, in slow-noise (fully correlated) limit, the FI of the CM monotonically increases with a larger $N$ while the information of AMR estimator is bounded by $1/c$. Extracting the FI comes at the cost of knowing the covariance matrix of the correlated noise and performing MLE to estimate the parameter of interest. Due to the post-selection in WVA, the slow-noise condition can be further divided into two cases according to whether the noise of successfully post-selected data is correlated or not. In slow-noise limit 1, post-selection in WVA eliminates the correlation of noise. Thus, AMR estimator is effective to achieve larger metrological information than that of the CM. In slow-noise limit 2, though post-selected data are still correlated, the amplification effect of the parameter is beneficial to mitigate the unknown off-set $c$. For a large $N$, the metrological information of AMR estimator $\mathcal{I}_{\text{WVA}}$ is $1/p_f$ times larger than $\mathcal{I}_{\text{CM}}$.

\subsection{5.2 Imaginary weak-value amplification against technical noise}

In the imaginary WVA, the parameter is measured in the Fourier conjugate space, which provides the potential to mitigate the technical noise. In 2012, Kedem demonstrated that a specific type of technical noise is beneficial to improve the SNR via imaginary WVA \cite{kedem2012using}. In 2014, Jordan et al. focused on the beam-deflection measurements in the presence of combined noise types, such as detector transverse jitter, angular beam jitter of the initial MS, and turbulence \cite{jordan2014technical}. Since both the quantum measurements and the AMR estimator are optimal to estimate the parameter in these cases, the SNR and FI metrics are equivalent to evaluate the precision. In this subsection, we summarize the the metrological advantages of imaginary WVA in the measurement of beam deflection using FI metric. 

\begin{figure*}[t]
	\centering
	\includegraphics[width=0.75\textwidth]{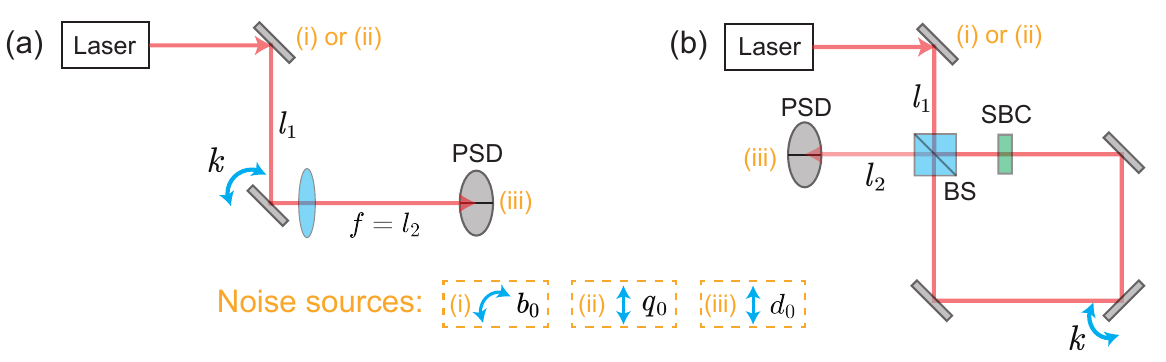}
	\caption{Comparison between the imaginary WVA and CM in the presence of dynamical noise. (a) Measurement of the mirror deflection $k$ in CM. (b) Measurement of the mirror deflection inside the Sagnac interferometer $k$ in the imaginary WVA. The noise (i) and (ii) is simulated by the random deflection and displacement $b_0$ and $q_0$ of the front mirror, respectively. The noise (iii) is modelled by the random lateral displacement $d_0$ of detector. BS, beam splitter; SBC, Soleil-Babinet compensator, PSD, position-sensitive detector.}
	\label{fig:im_WVA_noise}
\end{figure*}

Fig. \ref{fig:im_WVA_noise} (a) depicts the measurement of the mirror's deflection $k$ in CM. The implementation of the imaginary WVA to measure the mirror's deflection inside the Sagnac interferometer is shown in Fig. \ref{fig:im_WVA_noise} (b). To investigate the effects of dynamical noise in the CM and WVA, we summarize three cases: (i) the noise of the initial MS is modelled by the deflection $b_0$ of the first mirror, in which $b_0$ follows a Gaussian distribution with zero mean and variance $B_0^2$; (ii) the noise of the initial MS is modelled by the displacement $q_0$ of the first mirror, in which $q_0$ follows a Gaussian distribution with zero mean and variance $Q_0^2$; (iii) the jitter noise of the detectors is modelled by the lateral displacement $d_0$, which follows a Gaussian distribution with zero mean and variance $D_0^2$.

In case (i), the initial Gaussian MS $|\Phi\rangle$ evolves sequentially with the following operations: a random momentum kick $\hat{U}_{b_0} = \exp(ib_0\hat{x})$, free propagation over a distance $l_1$: $\hat{U}_{l_1} = \exp(-i\hat{p}^2l_1/2k_0)$, momentum kick $\hat{U}_{k} = \exp(ik\hat{x})$, quadratic phase modulation by a lens $\hat{U}_f = \exp(-ik_0 \hat{x}^2/2f)$, and free propagation over a distance $l_2$: $\hat{U}_{l2}$, leading to the final spatial distribution on the detector 
\begin{equation}
P_d(x,p_0) = |\langle x|\hat{U}_{l_2}\hat{U}_{f}\hat{U}_{k}\hat{U}_{l_1}\hat{U}_{p_0}|\Phi\rangle|^2,
\end{equation}
where $k_0$ is the wave number of the light. After averaging the random variable $b_0$, we obtain a Gaussian distribution with an average $\langle x \rangle = -kf/k_0$ and a variance $\langle (x-\langle x\rangle)^2\rangle = f^2[1/(2\sigma k_0)^2 + P_0^2/k_0^2]$. Thus, the FI about the parameter $k$ is given by
\begin{equation}
F_{\text{CM}} = \frac{4N\sigma^2}{1+(2\sigma B_0)^2},
\label{Eq:CM_noise1}
\end{equation}
where $N$ is the number of photons detected and $\sigma$ is the width of the Gaussian MS. Since the angular jitter noise $b_0$ directly adds to the signal deflection $k$, leading to the reduction of the FI by a factor $1/[1+(2\sigma B_0)^2]$. 

In the imaginary WVA, the QS is encoded on the which-way information of the Sagnac interferometer with $|0\rangle$ ($|1\rangle$) referring to the clockwise (counterclockwise) photonic states. Given the observable of the QS $\hat{W} = |0\rangle\langle 0| - |1\rangle\langle 1|$, the momentum kick operator $\hat{U}_k$ in the CM is substituted by the weak coupling unitary process of the imaginary WVA represented as $\hat{U}_{kW} = \exp(ik\hat{W}\hat{x})$. The pre- and post-selected states are prepared as $|\psi_i\rangle = (|0\rangle+i|1\rangle)/\sqrt{2}$ and $|\psi_f\rangle = (|0\rangle-i|1\rangle)/\sqrt{2}$, respectively. The relative phase shift that is induced by a SBC inside the Sagnac interferometer is represented by the operator $\hat{U}_\phi = \exp(i\phi \hat{W}/2)$. Consequently, the resulting probability distribution at the detectors is given by
\begin{equation}
P_{\text{WVA}}(x,q) = |\langle x|\langle \psi_f| \hat{U}_{l_2}\hat{U}_{kW}\hat{U}_{\phi}\hat{U}_{l_1}\hat{U}_{b_0} |\psi_i\rangle|\Phi\rangle|^2.
\end{equation}
This distribution reveals an average shift of the post-selected MS 
\begin{equation}
\langle x \rangle \approx \frac{4k\sigma^2}{\phi} + \frac{kl_1(l_1+l_2)}{k_0^2 \phi \sigma^2} + \frac{b_0(l_1+l_2)}{k_0},
\label{Eq:Ave_shift_noise1}
\end{equation}
and the variance is given by
\begin{equation}
\langle (x-\langle x\rangle )^2\rangle = \sigma^2 + \Big(\frac{l_1+l_2}{2k_0 \sigma}\Big)^2.
\end{equation}
Assuming that the distribution can be approximated as a Gaussian with the aforementioned mean and variance, the FI about the parameter $k$ in the imaginary WVA is calculated as
\begin{equation}
F_{\text{WVA}} = \frac{4N\sigma^2}{1+(\frac{l_1+l_2}{2k_0\sigma^2})^2[1+ (2\sigma B_0)^2]}.
\label{Eq:WVA_noise1}
\end{equation}
From the average shift of the MS in Eq. \eqref{Eq:Ave_shift_noise1}, the angular jitter noise $p_0$ results in a small correction to the signal deflection due to the beam diffraction. Thus, compared to the FI of the CM in Eq. \eqref{Eq:CM_noise1}, the reduction of the FI caused by the noise $p_0$ is mitigated by a diffraction factor of $(l_1+l_2)/(2k_0\sigma^2)$ in the imaginary WVA. The factor is typically much smaller than 1, indicating the metrological advantages of the imaginary WVA over the CM in the presence of angular jitter noise.

In case (ii), a random displacement $q_0$ occurs at the first mirror leading to the initial MS $|\Phi (q-q_0)\rangle$ with the probability $\text{Pr}(q_0) = 1/[(2\pi)^{1/2}Q_0]\exp(-q_0^2/(2Q_0^2))$, where $Q_0^2$ is the variance of the noise. The interaction process of measuring the deflection is still $\hat{U}_{kW}$. The relative phase shift $\hat{U}_\phi$ can be combined to the pre- and post-selected states resulting in $|\psi_i\rangle = (|0\rangle+i|1\rangle)/\sqrt{2}$ and $|\psi_f\rangle = (e^{-i\phi/2}|0\rangle-ie^{i\phi/2}|1\rangle)/\sqrt{2}$. Thus, the imaginary weak value is given by $\langle \hat{A}\rangle_w = -i2/\phi$. After the interaction and post-selection, the post-selected MS before the detectors is given by 
\begin{equation}
|\Phi\rangle_f = \mathcal{N}_{q_0}(\frac{2\sigma^2}{\pi})^{\frac{1}{4}} \exp[-\sigma^2 (p - q_0)^2 - 2kp/\phi],
\end{equation}
where $\mathcal{N}_{q_0}$ is the normalization factor. Measurements of the observable $\hat{P} = |p\rangle\langle p|$ in $|\Phi\rangle_f$ yield the average shift and the variance 
\begin{eqnarray}
\langle \hat{P}\rangle_f &=& q_0 - \frac{k}{\phi \sigma^2} {}\nonumber\\
\langle \hat{P}^2\rangle_f &=& \frac{1}{4\sigma^2} + (q_0 - \frac{k}{\phi \sigma^2} )^2.
\end{eqnarray}
The success probability of post-selection is given by
\begin{equation}
\text{Pr}(|\Phi\rangle_f |q_0) = |\langle \psi_f|\psi_i\rangle|^2 \mathcal{N}_{q_0}^{-2}.
\end{equation}
When the post-selected MS $|\Phi\rangle_f$ is successfully obtained, the displacement noise follows the distribution:
\begin{eqnarray}
\text{Pr}(Q_0||\Phi\rangle_f) &=& \frac{\text{Pr}(q_0)\text{Pr}(|\Phi\rangle|q_0)}{\text{Pr}(|\Phi\rangle)}  {}\nonumber\\
& = & \frac{1}{\sqrt{2\pi}Q_0} \exp[-\frac{q_0 + 2kQ_0^2/\phi}{2Q_0^2}].
\end{eqnarray}
By averaging the noise term $q_0$, the final average shift and the variance of the post-selected MS on the detectors are given by
\begin{eqnarray}
\overline{\langle \hat{P}\rangle_f} &=& \frac{k}{2\phi}(\frac{1}{(2\sigma)^2} + 2Q_0^2)  {}\nonumber\\
\overline{\langle \hat{P}^2\rangle_f} - \overline{\langle \hat{P}\rangle_f}^2 &=& \frac{1}{\sigma^2} + Q_0^2.
\end{eqnarray}
Therefore, the FI about the parameter $k$ in WVA is given by
\begin{equation}
F_{\text{WVA}} = 4N (\sigma^2 + Q_0^2).
\end{equation}
We can find that the incoherent spread in the Fourier space of the initial MS is beneficial for the imaginary WVA to improve the precision.

In case (iii), the transverse position of detectors shifts with a random variable $d_0$, which subjects to zeros-mean Gaussian distribution with the variance $D_0^2$. In the CM, the transverse detector jitter simply increases the average beam waist to $\sqrt{\sigma^2 + D_0^2}$, leading to the new FI
\begin{equation}
F_{\text{CM}} = \frac{N(f/k_0)^2}{\sigma_f^2 + D_0^2} = \frac{4N\sigma^2}{1 + (\frac{2k_0\sigma D_0}{f})^2},
\end{equation}
where $\sigma_f$ is the focused beam waist given by $\sigma_f = f/(2k_0\sigma)$. The result indicates that the noise term $(\frac{2k_0\sigma D_0}{f})^2$ in the denominator can be suppressed by a large focal length $f\ge 2k_0\sigma D_0$. This is equivalent to generate a large focal spot $\sigma_f$ as well as induce a large displacement $fk/k_0$, which intuitively mitigates the reduction of FI due to the detector jitter.

In the imaginary WVA, the detector jitter results in the new FI of
\begin{equation}
F_{\text{WVA}} = \frac{4N\sigma^2}{1 + D_0^2/\sigma^2}.
\end{equation}
Therefore, a large initial beam waist $\sigma$ not only increases the original FI but also helps to suppresses noise. Therefore, the practical advantages of WVA over the CM mainly lie in two aspects: (i) The widely used PSD such as split detectors have a separation between two photoelectric detectors, which establishes a minimum beam diameter and consequently restricts its propagation through horizontal turbulence in CM. (ii) In WVA, the PSD can be placed immediately after the BS to shorten the optical path length, which mitigates the possible noise caused by the air turbulence.

\subsection{5.3 Weak-value amplification with imperfect detectors}
In the preceding two subsections, we have showcased the metrological advantages of WVA in the presence of dynamical noise. Furthermore, the imperfections of photodetectors also play a crucial role in influencing the measurement precision in an optical system. Pixelation, classical electrical noise and saturation effect are common imperfections found in an arrayed photodetectors such as change-coupled devices (CCDs) and complementary metal oxide semiconductor (CMOS). In the following, we focus on discussing whether WVA can provide advantages over the CM in the presence of these imperfections.

\begin{figure*}[t]
	\centering
	\includegraphics[width=0.9\textwidth]{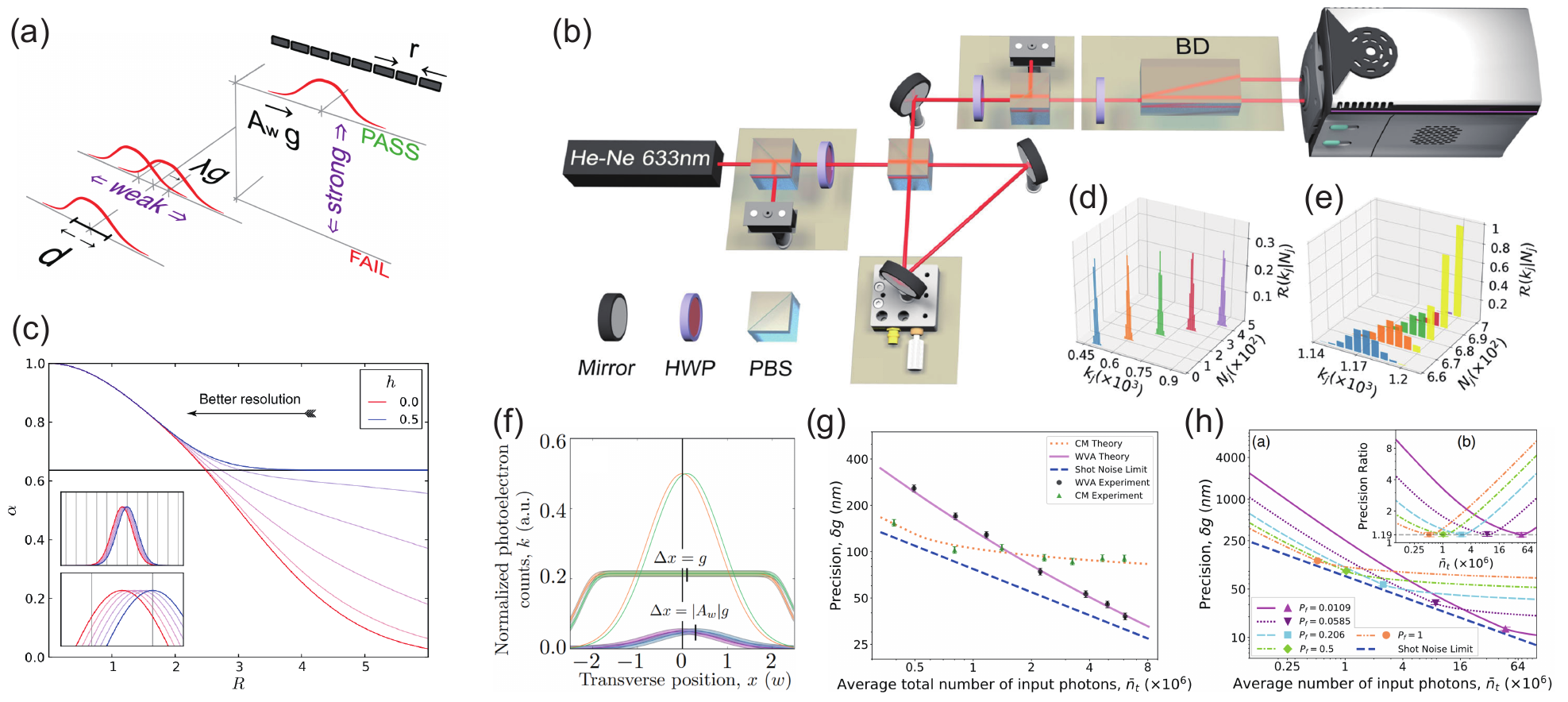}
	\caption{WVA with imperfect detectors. (a) The basic procedure of the standard WVA scheme and the post-selected MS is detected by arrayed detectors with the pixel size $r$ \cite{knee2014amplification}. (b) The experimental setup to implement the real WVA scheme to measure a small displacement of a mirror inside the Sagnac interferometer \cite{xu2020approaching}. (c) Numerically obtained relationships between relative information $\alpha$ and inverse resolution $R$ \cite{knee2014amplification}. (d) and (e) illustrate the response matrix of the CCD before and after the saturation, respectively \cite{xu2020approaching}. (f) Comparison of the beam profiles between the CM and the WVA in the presence of detector saturation \cite{harris2017weak}. (g) Comparison between the precision of the CM and that of the WVA \cite{xu2020approaching}. (h) The precision of the real WVA with different success probability of post-selection \cite{xu2020approaching}.}
	\label{fig:saturation}
\end{figure*}

In 2014, George et al. conducted the initial analysis on how the pixelation affects the ultimate precision to estimate a small lateral displacement of a Gaussian beam \cite{knee2014amplification}. As depicted in Fig. \ref{fig:saturation} (a), the pixelation of arrayed photodetectors transforms the probability density function $P(q)$ into a discrete probability distribution, given by
\begin{equation}
\text{Pr}(n) = \int_{(n-1/2)r}^{(n+1/2)r} P(q) dx,
\end{equation}
where $r$ represents the width of the pixel. The pixel number $n$ is defined as the Integer part of $q/r$, denoted as $n = \lfloor q/r \rceil$. Generally, we assume that the wave function is shifted by $\nu g$, leading to the final probability distribution $P(q-\nu g)$. When the pixel size tends towards infinitely small, the coefficient $\nu$ can be separated from the pixelation effect in the FI about the parameter $g$
\begin{equation}
F[Pr(n^\prime)] = \nu^2 F[Pr(n)],
\end{equation}
where $n^\prime = \lfloor (q-\nu g)/r \rceil$. However, as the pixel size increases, the misalignment between the centroid of $\text{Pr}(n^\prime)$ and the pixel boundaries become significant. This relative alignment is described by $h = \nu g -\mu \text{mod} r$ with $\mu$ a controllable quantity. Consequently, the FI about $g$ with pixelated detectors can be expressed as
\begin{equation}
F_g[\text{Pr}(\lfloor (q-\nu g)/r \rceil)] = \nu^2 F_g[(\lfloor (q-h)/r \rceil)]
\end{equation}
As the relation between $h$ and $\nu$ is somewhat arbitrary, a larger $\nu$ does not necessarily imply a better choice of $h$ to increase $F_g$. In comparing the FI between the real WVA and the CM, we take the identical $h$, resulting in
\begin{equation}
\frac{p_f F_g[\text{Pr}(\lfloor (q-\text{Re}(A_w) g)/r \rceil)]}{F_g[\text{Pr}(\lfloor (q- \lambda_{\max} g)/r \rceil)]} = \frac{Re(\langle \psi_f|\hat{A}|\psi_i\rangle)^2}{\lambda_{\max}^2} \le 1.
\end{equation}
Therefore, the pixelation effect is cancelled for both the real WVA and the CM. The amplification effect cannot compensate for the reduction of information due to pixelation. Similarly, the ratio of FI between the imaginary WVA and CM is given by
\begin{equation}
\frac{p_f F_g[\text{Pr}(\lfloor (p-\text{Im}(A_w) g)/r \rceil)]}{F_g[\text{Pr}(\lfloor (q- \lambda_{\max} g)/r \rceil)]} = \frac{\alpha_p}{\alpha_q},
\end{equation}
where the coefficients $\alpha_p$ and $\alpha_q$ represent the ratio of FI between the pixlated detection and the ideal detection. The numerical relationship between the information ratio $\alpha$ and $h$ is illustrated in Fig. \ref{fig:saturation} (c). When $R$ approaches $0$, the information ratio tends towards $1$ which is quite robust against the misalignment $h$. However, as $R$ becomes larger, $h=0.5$ is the best choice for the detection. Since $\alpha_p$ and $\alpha_q$ approximately approach unity to the first order of the relative resolution $R = r/\Delta q$, we have $\alpha_p/\alpha_q\approx 1$. Thus, the imaginary WVA cannot offer metrological advantages over the CM. This conclusion appears to contradict the findings of Brunner et al., where the imaginary WVA combined with frequency-domain detection has the potential to outperform the CM in measuring small longitudinal phase shifts. In Brunner's paper, the main experimental constraints are the alignment errors of the equipment and the resolution limit of the detectors. We need to emphasize that the resolution limit is not equivalent to the pixel size of the arrayed detectors. It can be interpreted as the ultimate sensitivity of detecting the shift, incorporating the effect of pixelation and other photo-electrical noise.

When $R$ tends to infinity, it corresponds to the case of split detectors, which is popular in monitoring the lateral displacement of a light beam. Once the optimal misalignment parameter $h = 0.5$ is taken, only about a third information is lost during the detection. In 2015, Knee et. al. have studied both the SNR and FI metrics in estimating a Gaussian beam displacement with a split detector \cite{knee2015fisher}. The use of the split detectors is optimized to obtain the most FI. The study implies that the amplification effect cannot bring improvement of FI with split detection in the absence of other technical noise, the real WVA is considered to be strictly worse than the CM. Nevertheless, the split detection may maintain the robustness of WVA scheme to certain technical noise and therefore is an effective detection method in WVA.

By exploiting noise-robust measurement schemes, we can effectively improve the precision of estimating an parameter in the practical situations. However, the precision of estimation is ultimately limited by the SNL due to the quantum fluctuations of light in the optical metrology. Although the use of quantum resources, such as entangled and squeezed states, can push the measurement precision to the HL, the preparation and control of these quantum states are currently complicated, limiting their practical applications. An alternative approach is to increase the average photon number of coherent states, which improves the measurement precision along the SNL. In optical systems, a frequent constraint of this approach is the saturation of photodetectors. The distortion of detecting profiles caused by saturation may lead to the deviation of the precision improvement from the SNL. Therefore, how to enhance the dynamic range of the measurement systems and mitigate the information loss due to the detector saturation is crucial in the practical metrology tasks. In section 3.3, theoretical analysis has indicated that a small fraction of post-selected photons in WVA are able to carry as much metrological information as almost all the input photons. This provides WVA a potential to avoid the loss of the information due to the detector saturation and equivalently extend the dynamic range of detectors.

In 2017, Harris et. al. proposed the theoretical framework to compare the measurement precision of CM and WVA by carefully considering the pixelated, digitized, and noisy response detector with the saturation threshold \cite{harris2017weak}. In their model, the beam profiles of CM and WVA are illustrated in Fig. \ref{fig:saturation} (f). Their numerical results indicated that WVA shows metrological advantages over the CM with the saturation effect in combination with the pixel or digitization noise. In 2020, Xu et al. optimized the WVA scheme by maximizing the FI over the probability of post-selection to fully investigate the capability of the standard WVA in the resistance against detector saturation \cite{xu2020approaching}. As introduced in section 3.3, the precision of the real WVA scheme is approximately the same as that of CM and maintains almost unchanged along with the reduction of the post-selection probability within a certain range. Given the number of input photons $N_j$ and the readout $k_j$ on the $j$th pixel of the CCD, the response matrix $\mathcal{k_j|N_j}$ of the photodetectors are carefully calibrated and illustrated in Figs. \ref{fig:saturation} (d) and (e) before and after the saturation, respectively. The readout of the pixel is bounded by the saturation threshold $k_s$. Assume that the average number of photons inputting on the $j$th pixel is $\bar{n}_j$, the probability distribution of readouts on the $j$th pixel is given by
\begin{equation}
P(k_j|g) = \sum_{N_j}\mathcal{R}(k_j|N_j)P(N_j|\eta \bar{n}_j,g),
\end{equation}
where $\eta$ is the detection efficiency of the detector and $g$ is the coupling strength to be estimated. This expression is applicable to both the WVA and the CM, with the difference lying in the $\bar{n}_j$. If the input light is in coherent state and the statistical distribution of photon numbers subjects to Poisson distribution, the FI about the parameter $g$ can be simplified as
\begin{equation}
F(g) = \sum_{j}\frac{\eta}{\bar{n}_j}\left(\frac{d\bar{n}_j}{dg}\right)^2 \Gamma(\mathcal{R},\bar{n}_j),
\end{equation}
in which $\Gamma(\mathcal{R},\bar{n}_j)$ can be interpreted as an equivalent SNR at $j$th pixel. As illustrated in Fig. \ref{fig:saturation} (b), the displacement of mirror inside the Sagnac interferometer is measured with the real WVA and the CM, respectively. The parameter is estimated with the MLE. The experimental results in Fig. \ref{fig:saturation} (g) has demonstrated a distinct advantage of WVA over the CM in the presence of detector saturation. Moreover, the optimal precision WVA over a range of success probabilities of post-selection $p_f$ allows to minimize the overall detector imperfections and maximize the precision. As Fig. \ref{fig:saturation} (h) shows, the variances of the optimal precision WVA maintains about 1.19 times larger than the SNL when the input photons saturates more and more certain pixels.

\section{6. Various modifications of weak-value amplification schemes}

\subsection{6.1 Inverse weak-value amplification}
\begin{figure*}[t]
	\centering
	\includegraphics[width=0.8\textwidth]{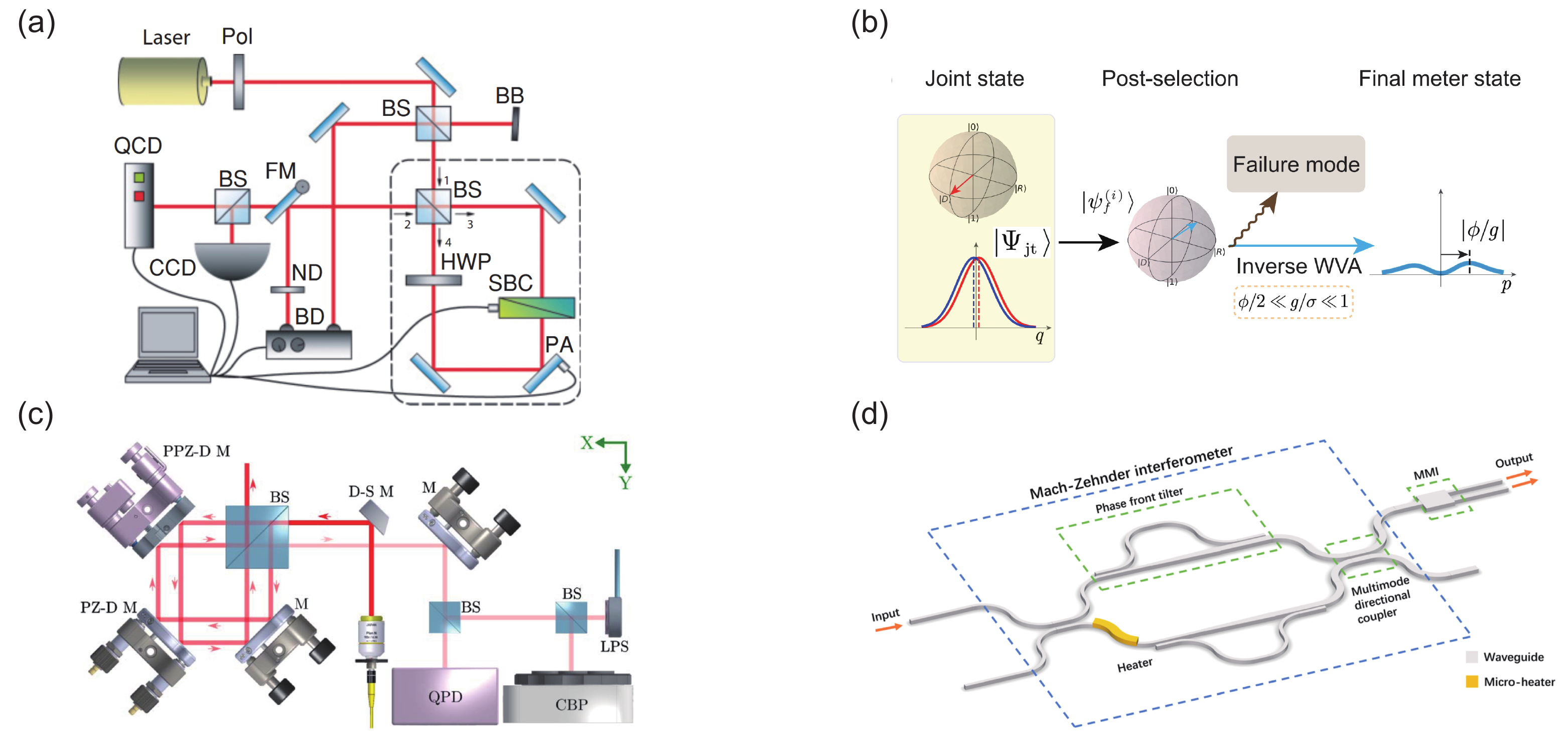}
	\caption{(a) The relative phase caused by half-wave plate (HWP) and Soleil-Babinet compensator (SBC) between two paths inside the Sagnac interferometer is inversely proportional to the transverse momentum shift and can be precisely measured by monitoring the beam profile in the dark port \cite{starling2010continuous}. (b) Schematic diagram for the inverse WVA. (c) The relative phase due to the mirror's tilt in a modified Sagnac interferometer is amplified by the tilt of another mirror in the perpendicular direction \cite{martinez2017ultrasensitive}. (d) The inverse WVA is realized with phase front tilters in an on-chip Mach-Zehnder interferometer \cite{song2021enhanced}. }
\label{inverse_WVA}
\end{figure*}

The inverse WVA was first experimentally studied by Starling et al. to amplify the relative phase between two optical paths with the experimental setup depicted in Fig. \ref{inverse_WVA} (a) \cite{starling2010continuous}. The post-selection results in a transverse bimodal pattern at the dark port and the small phase can be inferred from the mean of the distribution. In 2012, Kofman et al. provided a theoretical description of the inverse WVA from the perspective of quantum theory \cite{kofman2012nonperturbative}. In Fig. \ref{inverse_WVA} (b), we demonstrate the inverse WVA scheme from the joint state $|\Psi_{jt}\rangle$ which originates from the interaction between the QS pre-selected by the state $|\psi_i\rangle = (|0\rangle+|1\rangle)/\sqrt{2}$ and a Gaussian MS $|\Phi\rangle$ in Eq. \eqref{eq:initial_MS} under the Hamiltonian $\hat{H} = g\delta(t-t_0)\hat{A}\hat{P}$. It is worth noting that in the inverse WVA, the coupling strength $g$ is set to measure the unknown parameters $\theta_I$ and $phi_I$ encoded in the QS. For the instance, the post-selected state involves the unknown parameter, given by $|\psi_f\rangle = \cos(\pi/4 - \theta_I/2)|0\rangle -\sin(\pi/4 - \theta_I/2)e^{i \phi_I}|1\rangle$. The post-selection on the QS leads to the final MS $|\Phi_f\rangle$. The inverse WVA works under the condition that the overlap between the pre- and post-selection is much smaller than the coupling strength, i.e., $|\langle \psi_f|\psi_i\rangle|\ll g/\sigma \ll 1$. Thus, both parameters in $|\psi_f\rangle$, namely $\theta_I$ and $\phi_I$ tend towards 0. The measurement results of the final MS exhibit a bimodal pattern instead of the Gaussian distribution in standard WVA. For simplicity, we consider two cases where the weak value $\langle \hat{A}\rangle_w = \cot (\theta_I/2) - i\cot(\phi_I/2)$ is purely real ($\phi_I=0$) or imaginary ($\theta_I=0$). According to Eq. \eqref{eq:wva_disp_exact}, the approximate average shifts in the observables $\hat{P}$ and its conjugate $\hat{Q}$ in the final MS are given by
\begin{equation}
\langle \hat{Q}\rangle_f \approx \frac{4\sigma^2}{g[\langle \hat{A}\rangle_w]_r} = \frac{2\theta_I\sigma^2}{g}
\end{equation}
for the real weak value $[\langle \hat{A}\rangle_w]_r \approx 2/\theta_I $ and 
\begin{equation}
\langle \hat{Q}\rangle_f \approx \frac{2}{g[\langle \hat{A}\rangle_w]_i} = -\frac{\phi_I}{g}
\end{equation}
for the imaginary weak value $[\langle \hat{A}\rangle_w]_r \approx -2i/\phi_I $. Correspondingly, the success probabilities of post-selection for the real and imaginary weak values are given by $p_f = g^2/(4\sigma^2)$ and $p_f = \phi_I^2/4$, respectively. It can also be proved that the inverse WVA is capable of recovering the complete FI about the parameter $\theta_I$ or $\phi_I$ through observing the post-selected MS. This explains why the sensitivity of phase measurement in \cite{starling2010continuous} is similar to balanced homodyne detectors through monitoring only the dark port of the interferometer.

In 2017, Mart\'{i}nez-Rinc\'{o}n et al. applied the inverse WVA to measure the tilt of a mirror in a modified Sagnac interferometer, as shown in Fig. \ref{inverse_WVA} (c) \cite{martinez2017ultrasensitive}. The mirror's tilt induces a phase difference between the counter-propagating light, and the tilt in the orthogonal plane couples the QS and the MS to amplify the phase. The experiment achieved a shot-noise-limited sensitivity of 56 $\text{frad}/\sqrt{\text{Hz}}$ using 1.2 mW continuous-wave laser, displaying excellent performance at low frequencies. In 2018, Lyons et al. demonstrated that the inverse WVA exhibits similar robustness to specific noise sources, such as additive Gaussian white noise and angular jitter noise, compared to standard WVA \cite{lyons2018noise}. In 2021, the inverse WVA was implemented on an integrated photonic platform utilizing a multi-mode interferometer, as shown in Fig. \ref{inverse_WVA} (d) \cite{song2021enhanced}. This setup showed a 7 dB enhancement in phase measurement compared to a standard Mach-Zehnder interferometer.

\subsection{6.2 Almost-balanced weak-value amplification}

\begin{figure*}[t]
	\centering
	\includegraphics[width=0.9\textwidth]{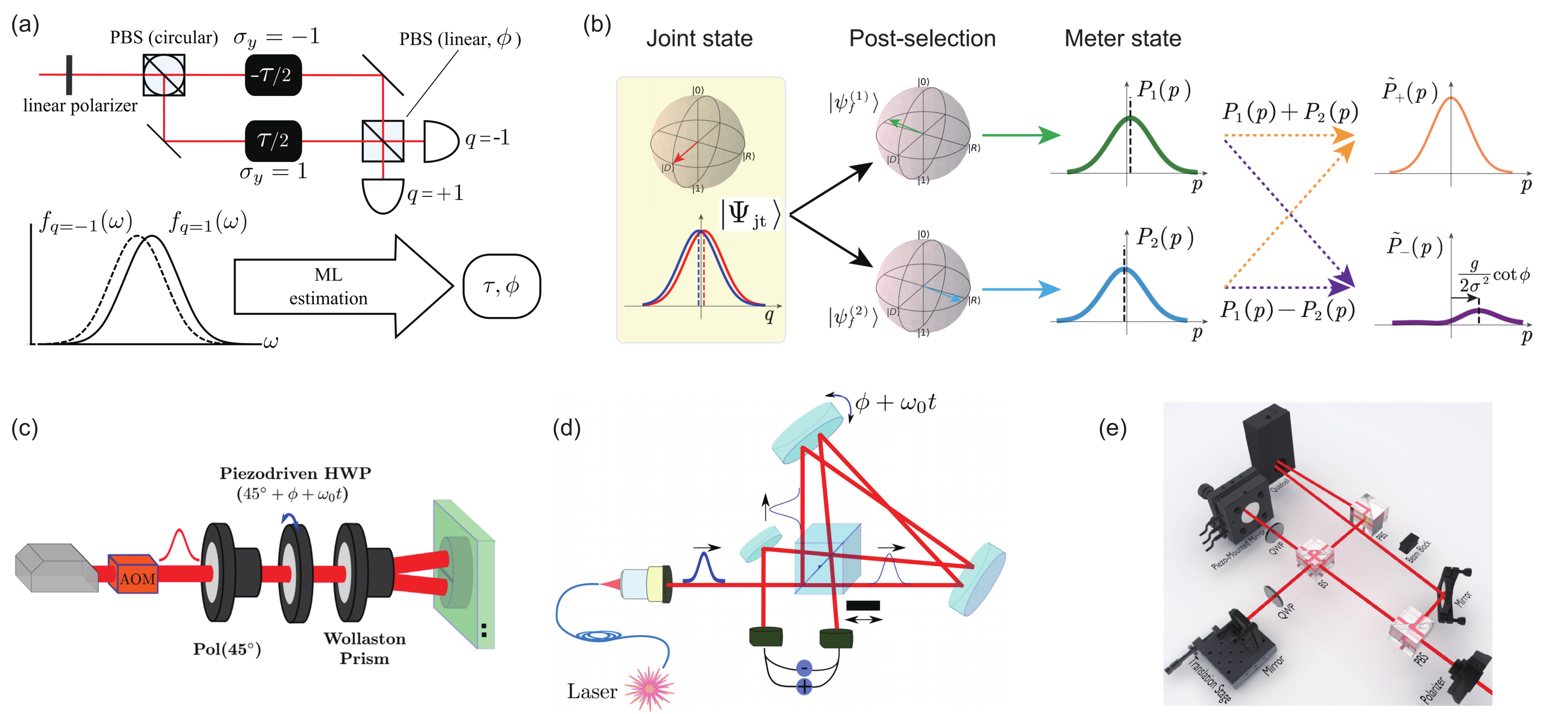}
	\caption{(a) The schematic of measuring ultra-small time delays of light by joint weak measurement \cite{strubi2013measuring}. (b) The schematic diagram of the almost-balanced WVA. (c) Experimental setup for measuring small angular velocities $\omega_0$ of the half-wave plate (HWP) without post-selection \cite{martinez2016can}. (d) Experimental setup for measuring the rotation of a mirror via ABWVA \cite{liu2017anomalous}. (e) Experimental setup for measuring linear velocities of a piezo-mounted mirror \cite{martinez2017practical}.}
	\label{fig:AB_WVA}
\end{figure*}

In section 3.2, we have discussed the distribution of metrological information during the post-selection process in WVA. Both the successful and failure modes of the MS generally contain the FI of the coupling strength $g$. The optimization of the standard WVA aims to concentrate all information about $g$ into the successful mode, which exhibits distinct advantages in suppressing certain technical noise and avoiding saturation. However, the significant reduction in signal intensity may lead to a terrible SNR due to the stray light and electrical noise of detectors. To resolve these problems, Strubi and Bruder proposed a joint measurement protocol that exploits the full information of the correlations between the QS and the MS \cite{strubi2013measuring}. The joint scheme is illustrated in Fig. \ref{fig:AB_WVA} (a) to measure a small relative time delay $\tau$ between two paths of a Mach-Zehnder interferometer. The QS (path DOF) and the MS (spectrum DOF $p_0(\omega)$) interact under the Hamiltonian $\hat{H} = -\tau \hat{\sigma}_y \Omega$, where the Pauli operator $\hat{\sigma}_y$ is the observable of the QS and $\hat{\Omega}$ is the angular frequency observable of the MS with the eigenvalue $\omega$. The QS is pre-selected by $|\psi_i\rangle = (|-\rangle + |+\rangle)/\sqrt{2}$ and the post-selection is performed with two orthogonal states, written as $|\psi_f^{(q)}\rangle = (|-\rangle + qe^{i\phi}|+\rangle)/\sqrt{2}$ with $q=-1$ or $1$. Subsequently, the post-selected MSs are individually detected in the frequency domain. Taking into account the Gaussian fluctuations of the alignment parameter $\phi$ subject to the noise kernel
\begin{equation}
\xi (\phi,\phi^\prime) = \frac{1}{\sqrt{2\pi}\epsilon} \exp\Big[-\frac{(\phi-\phi^\prime)^2}{2\epsilon^2} \Big],
\end{equation}
the probability distribution on detectors ($q = 1,-1$) is given by
\begin{equation}
P_q(\omega;\tau, \phi) = \frac{1}{2}P_0(\omega) [1+q\exp(-\epsilon^2/2)\cos(\phi-\omega \tau)]. 
\end{equation}
Based on the joint probabilities, both the coupling parameter $\tau$ and the parameter $\phi$ of the experimental setup can be estimated by maximizing the log-likelihood function $\tau_{est}, \phi_{est} = \max l(\tau, \phi) = \max \sum_{q=\pm 1}\int d\omega f_q(\omega) \log P_q(\omega; \tau, \phi)$. Considering that there is a zero-mean Gaussian white noise with the variance $\Omega^2$ for the frequency detection, the approximate estimation of $\tau$ in the presence of the fluctuation $\epsilon$ and detection noise $\Omega$ is
\begin{equation}
\tau = \tau_0 \Big[1+\frac{1}{2}\big( \frac{\epsilon}{\sin \phi}\big)^2 + \frac{1}{2}\big(\frac{\Omega}{\Delta \omega} \big)^2    \Big],
\end{equation}
where $\tau_0$ denotes the estimated $\tau$ without noise. The balanced post-selection, i.e., $\sin \phi \approx 1$ minimizes the effect of fluctuations. The detector noise $\omega$ can be effectively reduced with a wide frequency spread $\Delta \omega$, which coincides with the analysis in section 4.3. Compared to the standard WVA, the joint weak measurement scheme has the advantages in removing the systematic errors and the fluctuations in optical alignment during the parameter estimation. In 2016, Fang et al. developed a general theory for the arbitrary post-selection in weak measurement, which includes the standard WVA and the joint weak measurement as two special cases \cite{fang2016ultra}. With the experimental setup depicted in Fig. \ref{fig:AB_WVA} (c), they demonstrated the significant improvement in precision of joint weak measurement over the standard WVA when measuring ultra-small time delays. The joint weak measurement exhibited robustness against misalignment errors and imperfections in wavelength-dependent optical components.

Mart\'{i}nez-Rinc\'{o}n et al. further extended the joint weak measurement protocol and obtained a WVA-like response by subtracting the two post-selected readouts of detectors \cite{martinez2016can}. In Fig. \ref{fig:AB_WVA} (d), they employed polarization DOF and Gaussian temporal pulse as the QS and the MS, respectively, to measure the angular velocity of the rotating HWP. The experimental results confirmed that compared to the conventional measurement schemes, the joint weak measurement achieved improved sensitivity and precision. This technique was termed almost-balanced WVA (ABWVA) since the amplification effect stemmed from the differential signals of two almost-balanced post-selected results. Fig. \ref{fig:AB_WVA} (b) illustrates the basic procedure to implement the ABWVA. We start from the joint state $|\Psi_{jt}\rangle$ after the interaction between the pre-selected state $|\psi_i\rangle = (|0\rangle +|1\rangle)/\sqrt{2}$ and the Gaussian MS $|\Phi\rangle$ under the Hamiltonian $\hat{H} = g\hat{A}\hat{P}\delta (t-t_0)$. Two orthogonal states $|\psi_f^{(1,2)}\rangle = (|0\rangle \pm ie^{i\epsilon}|1\rangle)/\sqrt{2}$ perform post-selection on the QS, leading to two final MSs. Two probability distributions $P_{1,2}(p;g) = [1\pm \sin(\epsilon+2gp)]P_0(p)/2$ are obtained by measuring the final MSs, where $P_0(p) = |\Phi(q)|^2$. The distribution of the initial MS $P_0(p)$ can be recovered by $\tilde{P}_+(p) = P_1+P_2 = P_0(p)$, while the difference of the distribution gives rise to
\begin{equation}
\tilde{P}_-(p;g) = P_2-P_1\approx \sin (\epsilon) P(p - \frac{g}{2\sigma^2}\cot \epsilon).
\end{equation}
In ABWVA, the signal strength of $\tilde{P}_-(p;g)$ is linearly proportional to $\sin (\epsilon)\approx \epsilon$, which is improved by a factor of $4/\epsilon$ compared to the standard WVA. Thus, the ABWVA allows for larger amplification (smaller $\epsilon$) in practical situations limited by technical noise than the standard WVA protocol. This advantage has been experimentally demonstrated by Liu et al. The experimental setup is shown in Fig. \ref{fig:AB_WVA} (d). They have measured the angular velocity of a mirror $\omega_0$ inside the interferemeter using both the standard WVA and ABWVA \cite{liu2017anomalous}. The minimal detectable value of the parameter $\phi$ is $\sim 4 \mu \text{rad}$ and $\sim 83 \text{nrad}$ for the standard WVA and ABWVA, respectively. The experimental results indicate that the maximum amplification factors of ABWVA is 24 times larger than the optimal standard WVA. In Fig. \ref{fig:AB_WVA} (e), Mart\'{i}nez-Rinc\'{o}n et al. demonstrated the practical advantages of ABWVA over the standard WVA in larger SNR and longer integration time due to slow drifts. The experiment has reported an improvement of velocity measurements with the sensitivity of 60 fm/s using ABWVA \cite{martinez2017practical}. Importantly, the idea of balanced detection in the ABWVA protocol can also be combined with other generalized forms of WVA schemes to further improve its performance. The similar experimental results are obtained in \cite{luo2020anomalous}.

\subsection{6.3 Biased weak-value amplification}

\begin{figure*}[t]
	\centering
	\includegraphics[width=0.9\textwidth]{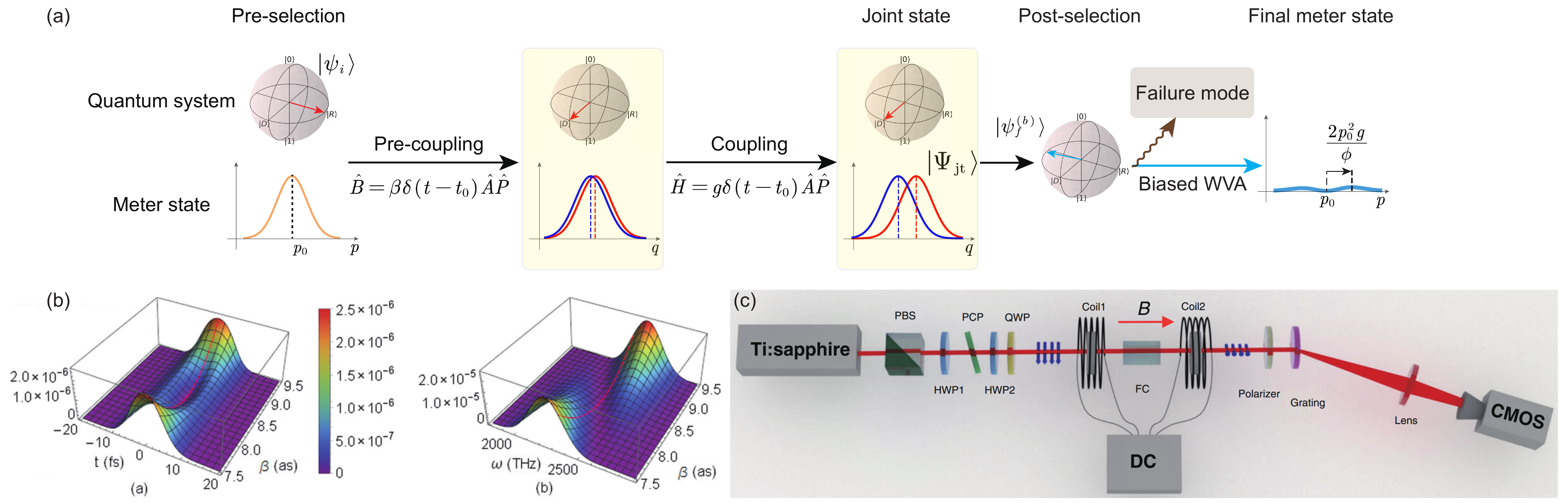}
	\caption{(a) The basic procedure for biased WVA scheme. (b) The probability density of the post-selected MS in the CDI regime \cite{zhang2016ultrasensitive}. (c) The experimental setup for the biased WVA to measure the magnetic strength \cite{yin2021improving}.}
	\label{Biased_WVA}
\end{figure*}

In 2016, Zhang et. al. proposed a biased WVA scheme by introducing a bias phase to the initial joint state of the QS and the MS before the standard WVA \cite{zhang2016ultrasensitive}. This bias allows for a much larger amplification factor than the standard WVA, with a correspondingly much lower success probability of post-selection. Fig. \ref{Biased_WVA} (a) illustrates the basic procedure of the biased WVA scheme. The initial joint state of the pre-selected state and the MS can be expressed as $|\psi_i\rangle\otimes |\Phi\rangle = (|0\rangle+i|1\rangle)/\sqrt{2} \otimes \int f(\omega) d\omega |\omega\rangle$, which evolves under a pre-coupling unitary process $\hat{U}_{pre} = \exp(-i\beta \hat{A}\hat{\Omega})$, where $\hat{A} = |0\rangle\langle 0| - |1\rangle\langle 1|$ and $\hat{\Omega}$ are the observables of the QS and MS, respectively. The pre-coupling leads to an entangled state, given by
\begin{equation}
|\Psi\rangle_{Ei} = \int d\omega \frac{1}{\sqrt{2}}f(\omega)[e^{i\omega\beta}|0\rangle + ie^{-i\omega\beta}|1\rangle]|\omega\rangle.
\end{equation}
The longitude phase change (LPC) $\omega \tau$ is encoded to the entangled initial state $|\Psi\rangle_{Ei}$ with the unitary transformation $\hat{U}_{\tau} = \exp(-i\tau \hat{A}\hat{\Omega})$. Subsequently, the joint state is post-selected by $|\psi_f\rangle = (ie^{i\epsilon}|0\rangle + e^{-i\epsilon}|1\rangle)$ giving the final MS
\begin{equation}
|\Phi_f\rangle = \frac{i}{2\sqrt{p_f}} \int d\omega f(\omega)\Big\{ \exp\big[i\omega(\beta +\tau)-i\epsilon \big] -\exp\big[-i\omega(\beta +\tau)+i\epsilon \big] \Big\}|\omega\rangle
\end{equation}
with the success probability of post-selection
\begin{equation}
p_f = \frac{1}{2}\Big\{ 1-\exp(-\delta^2 \beta ^2)\cos\big[2(\omega_0 \beta +\omega_0 \tau - \epsilon)  \big]  \Big\}.
\end{equation}
The probability distribution of $|\Phi_f\rangle$ in the frequency domain is 
\begin{equation}
S(\omega) = \sin^2 \big[ \omega (\beta+\tau) -\epsilon \big]|f(\omega)|^2.
\end{equation}
Compared to the standard WVA ($\beta = 0$) with the probability distribution $S_{WVA}(\omega) \approx \epsilon^2 |f(\omega)|^2$, the extra bias phase induces the conjugate destructive interference (CDI) and reshapes the frequency spectrum with an ultra-small $\tau$ around $\beta_s$ satisfying $\omega_0 \beta_s -\epsilon \approx 0$. The CDI in both time and frequency domain is illustrated in Fig. \ref{Biased_WVA} (b). The spectrum shift $\Delta \omega = \int S(\omega)\omega d\omega/\int S(\omega)d\omega - \omega_0$ caused by the LPC can be approximated as
\begin{equation}
\Delta \omega \approx \frac{d \Delta \omega}{d\tau}|_{\beta_s}\tau \approx 2\frac{\omega_0^2}{\epsilon}\tau,
\end{equation}
with the corresponding success probability $p_f \approx \delta^2 \epsilon^2 /(2\omega_0^2)$. 

Since the central frequency of light $\omega_0$ is much larger than its frequency spread $\delta_\omega$, the amplification factor of the biased WVA $2\omega_0^2/\epsilon$ largely surpasses the standard WVA $2\delta_\omega^2/\epsilon$. When the decisive factor limiting measurement sensitivity is the spectrum resolution, represented by $\Delta \Omega$, the ultimate resolution limit of the standard WVA and BWVA is given by
\begin{eqnarray}
\tau_s &>& \frac{|\epsilon|\Delta \Omega}{\delta_\omega^2}, {}\nonumber\\
\tau_B &>& \frac{|\epsilon|\Delta \Omega}{2\omega_0^2},
\end{eqnarray}
respectively. It is evident that biased WVA achieves higher resolution of $\tau$. However, since the success probabilities of post-selection in both standard and biased WVA are inversely proportional to the corresponding amplification factors, these two schemes achieve the same ultimate precision with ideal setups. In 2020, Yin et al. exploited this feature of biased WVA to overcome the detector saturation effect, achieving higher precision than standard WVA \cite{yin2021improving}. In 2019, Huang et. al. further develop the framework of biased WVA by introducing the joint measurement scheme at the detection stage \cite{huang2019toward}. This new protocol, called dual WVA, significantly improves the post-selection probability compared to BWVA, thus acquiring a higher SNR in the presence of practical imperfections.

\subsection{6.4 Power-recycling weak value amplification}

Based on the preceding discussion, it is evident that the standard WVA approach, alongside other generalized WVA schemes, cannot surpass the SQL in terms of measurement precision or signal-to-noise ratio (SNR) due to the inherent trade-off between the amplification effect and the post-selection probabilities. However, WVA exhibits a unique capability of extracting the complete metrological information from  all the input photons by detecting only a minority of post-selected photons. This feature enables to improve the measurement precision through recycling the photons that fail the post-selection. 

\begin{figure*}[t]
	\centering
	\includegraphics[width=0.9\textwidth]{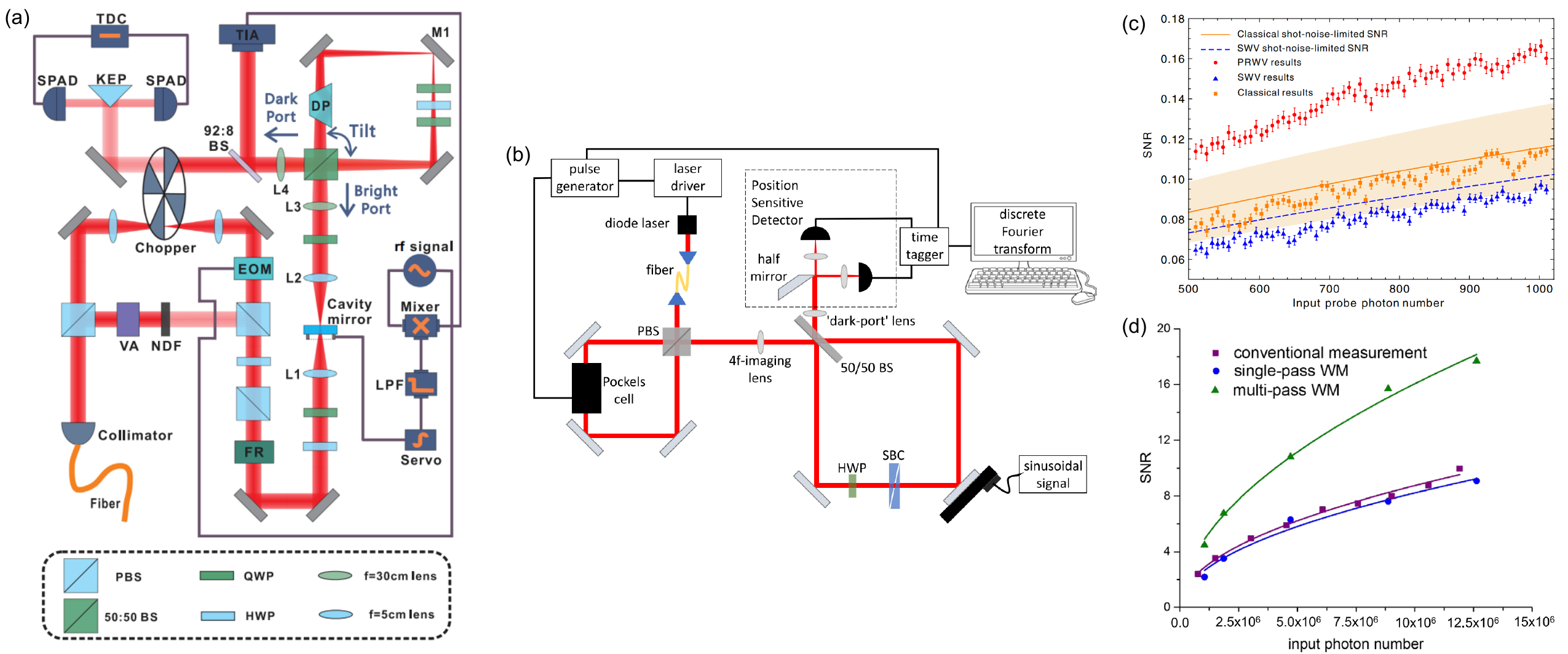}
	\caption{(a) The experimental setup for the power-recycling WVA using continuous light \cite{wang2016experimental}. (b) The experimental setup for the power-recycling WVA using pulsed light \cite{krafczyk2021enhanced}. (c) The experimental results for (a). (d) The experimental results for (b).}
	\label{fig:PRWVA}
\end{figure*}

In recent years, the power-recycling WVA schemes have been studied both theoretically and experimentally. In 2013, Dressel et al. proposed a theoretical scheme for recycling collimated and diverging beams by trapping the light pulse inside an interferometer with a Pockels cell and polarization operations \cite{dressel2013strengthening}. Their results demonstrated that the recycling of photons can improve the SNR of WVA measurements. In 2021, Krafczyk et al. experimentally realized this theoretical scheme using an experimental setup shown in Figure \ref{fig:PRWVA}(b) \cite{krafczyk2021enhanced}. The setup includes a combination of a Pockels cell and a loop of polarizing beam splitters to reinject the photons that fail the post-selection back into the WVA setup. The success probability of post-selection for one round of WVA is denoted as $p_f$ with the total photon number $N$. The remaining photons after $j$ rounds of recycling can be calculated as $N_j = N(1-p_f)^j$. Thus, a new total number of $\sum_{j=0}^\infty p_fN_j = N$ photons are detected without optical loss. Compared to the SNR of one-round WVA $\mathcal{R}_{\text{one}} = g\langle \hat{A}\rangle_w\sqrt{p_fN}/\sigma$, the SNR of all-round WVA is improved to $\mathcal{R}_{\text{one}} = g\langle \hat{A}\rangle_w\sqrt{N}/\sigma$. In the experiment, the success probability of post-selection is set to be $p_f = 0.03$ with the optical loss per round approximately $16\%$. About $81$ photons from the last round are recycled. When the beam-reshaping effects are considered, the recycling signal is $4.4\pm 0.2$ times the signal without recycling. As shown in Fig. \ref{fig:PRWVA} (d), the overall SNR of the recycling WVA is improved by a factor of 2.1, compared to the CM or the standard WVA.

In 2015, Lyons et al. conducted an investigation into the continuous wave power-recycling WVA scheme \cite{lyons2015power}. The key idea lies in constructing an optical resonant cavity using a partially transmitting mirror and incorporating it into the post-selection component of the WVA setup. The success probability of post-selection $p_f$ corresponds to the transmission rate of the second cavity mirror. To achieve impedance matching, the transmission rate of the first mirror must also be set to $p_f$, ensuring that the number of photons leaving the cavity matches the input. Under ideal conditions, the cavity enhances the light intensity with a gain factor of $G = 1/p_f$, resulting in an improvement of the SNR for the standard WVA by a factor of $1/\sqrt{p_f}$.  In 2016, Wang et al. experimentally implemented this power-recycling WVA proposal, as depicted in Fig. \ref{fig:PRWVA} (a) \cite{wang2016experimental}. The power gain coefficient within the cavity is given by $G = (1-r)/[1+(1-\beta)r - 2\sqrt{r(1-\beta)}]$, where $r\approx 0.7$ represents the reflectivity of the cavity mirror, and $\beta \approx 0.4$ denotes the loss factor of one transversal mode. Fig. \ref{fig:PRWVA} (b) illustrates that the SNR of the power-recycling WVA is enhanced by a factor 2.4 compared to the CM and the standard WVA.

\section{7. Conclusion and Outlook}

In this paper, we systematically review the fundamental principles of the standard WVA as well as its applications in optical metrology and sensing. The rigorous derivation of the WVA formalism extends the standard WVA to three regions, thus determining a maximum amplification effect in practical situations. This generalization also encompasses orthogonal pre- and post-selected states of the QS and various light field modes beyond the Gaussian distribution as the MS, which broadens the application scenarios of WVA for multi-parameter estimation tasks. To elucidate the metrological advantages of WVA, we summarize the previous evaluation of WVA using quantum parameter estimation theory. By integrating the quantum resources (e.g., entanglement or squeezing states) into the WVA scheme, one can surpass the SNL and even achieve the HL. When the interaction is implemented in the form of a non-linear Hamiltonian, WVA enables to achieve the Heisenberg-scaling precision with only classical resources. Due to the trade-off between the probability of successful post-selection and the amplification factor, the standard WVA does not offer advantages in precision compared to the CM schemes with ideal setups. However, WVA outperforms CM in terms of precision in the presence of certain experimental imperfections, such as dynamical noise, photo-electric noise and detector saturation. Furthermore, various modifications of WVA have been recently proposed to address the limitations of the standard WVA. The metrological advantages of WVA stem from the post-selection process, inspiring the exploration of other post-selected metrology schemes \cite{arvidsson2020quantum, lupu2022negative, arvidsson2023nonclassical}. In summary, after decades of development, WVA technology has become a mature noise-robust technique for high-precision quantum measurement and high-sensitivity quantum sensing. It is progressively gaining importance in quantum precision measurement tasks.

\begin{acknowledgments}
This work was supported by the National Key Research and Development Program of China (Grants No. 2023YFC2205802), National Natural Science Foundation of China (Grants No. 12305034, 12347104), Civil Aerospace Technology Research Project (D050105).
\end{acknowledgments}

\end{document}